%% file: main.tex
\documentclass[sigconf]{acmart}

\acmConference[EASE 2022]{EASE '22: Evaluation and Assessment in Software Engineering}{June 13–15, 2022}{Göteborg, Sweden}

\AtBeginDocument{%
  \providecommand\BibTeX{{%
    \normalfont B\kern-0.5em{\scshape i\kern-0.25em b}\kern-0.8em\TeX}}}

\usepackage{amsmath,amsfonts}
\usepackage{algorithmic}
\usepackage{graphicx}
\usepackage{textcomp}
\usepackage{xcolor}
\usepackage[utf8]{inputenc}
\usepackage{standalone}
\usepackage{graphicx}
\usepackage{pgfplots}
\usepackage{pgf-pie}
\usepackage{threeparttable}
\usepackage[listings,skins,breakable]{tcolorbox}
\usepgfplotslibrary{statistics}
\usepackage{multirow} 
\usepackage{url}
\usepackage{subcaption}
\pgfplotsset{compat=newest}
\usepackage{adjustbox}
\usepackage{hyperref}
\usepackage{array,booktabs}
\usepackage{listings}
\usepackage{xcolor,colortbl}
\definecolor{darkcerulean}{rgb}{0.0, 0.2, 0.4}
\definecolor{carnelian}{rgb}{0.7, 0.11, 0.11}
\definecolor{tropicalrainforest}{rgb}{0.0, 0.4, 0.3}
\usepackage{subfiles}  
\usepackage{textcomp}
\usepackage{siunitx}
\usepackage{csquotes}
\usepackage{makecell}
\usepackage{hyperref}
\newtcolorbox{quotebox}{colback=cyan!20,boxrule=0.4pt,colframe=black,fonttitle=\bfseries,top=2pt,bottom=2pt}
\usepackage{titlesec}
\usepackage{lipsum}
\DeclareMathAlphabet\mathbfcal{OMS}{cmsy}{b}{n}

\newcommand{\orbugtype}{performance or accuracy }
\newcommand{\bugtype}{performance and accuracy }
\newcommand{\Bugtype}{Performance and accuracy }
\newcommand{\BugType}{Performance and Accuracy }

\copyrightyear{2022} 
\acmYear{2022} 
\setcopyright{acmcopyright}\acmConference[EASE 2022]{The International
Conference on Evaluation and Assessment in Software Engineering 2022}{June
13--15, 2022}{Gothenburg, Sweden}
\acmBooktitle{The International Conference on Evaluation and Assessment in
Software Engineering 2022 (EASE 2022), June 13--15, 2022, Gothenburg, Sweden}
\acmPrice{15.00}
\acmDOI{10.1145/3530019.3530029}
\acmISBN{978-1-4503-9613-4/22/06}

\begin{document}

\title{On Reporting \BugType Bugs for Deep Learning Frameworks: An Exploratory Study from GitHub}

\author{Guoming Long}
\affiliation{%
  \institution{Department of Computer Science}
  \city{Loughborough University}
  \country{UK}}
\email{g.long@lboro.ac.uk}

\author{Tao Chen}
\authornote{Corresponding Author}
\affiliation{%
  \institution{Department of Computer Science}
  \city{Loughborough University}
  \country{UK}}
\email{t.t.chen@lboro.ac.uk}

\begin{abstract}
The tremendous success of Deep Learning (DL) has significantly boosted the number of open-sourced DL frameworks hosted on GitHub. Among others, \bugtype bugs are critical factors that affect the reputation of these DL frameworks, therefore understanding the practice of discovering and investigating them for DL is important. In this paper, we conduct an exploratory study on the nature of reporting \bugtype bugs for DL frameworks, aiming to improve our knowledge on this topic. Our study covers 10 most popular open-sourced DL frameworks on GitHub (e.g., TensorFlow, Keras, and PyTorch), based on which we sample 664 representative \bugtype bug reports out of a total population of 22,522. Through systematic analysis, we found that: (1) low speed is the primary reason that a performance bug related report is submitted but we see no consistent pattern for accuracy related ones; (2) most of the reports are about issues encountered in the training stage; (3) only a small proportion of the reports provide insufficient information to investigate; (4) the majority of the \bugtype bug reports (from 69\% to 100\%) are not related to the actual bug or regarded as unclassified; (5) around 50\% of the performance and accuracy bug reports, which indeed reveal bugs, are not resolved by direct patches. Deriving from the above, we discuss a set of actionable implications to the researchers, maintainers, and report submitters. To promote open science, the labeled dataset has been made publicly available at \texttt{\textcolor{blue}{\url{https://zenodo.org/record/6371676}}}.
\end{abstract}



\maketitle

\input{sec/introduction}
\input{sec/bg}
\input{sec/method}
\input{sec/data}
\input{sec/class}
\input{sec/result}

\input{sec/implication}
\input{sec/threat}

\input{sec/related}

\input{sec/con}

\small
\bibliographystyle{ACM-Reference-Format}
\bibliography{main}

\end{document}

%% file: sec/introduction.tex
\section{Introduction}
\label{sec:intro}

Deep learning (DL), which is a kind of machine intelligence algorithms that mimics the workings of the human brain in processing data~\cite{goodfellow2016deep}, has been gaining momentum in both academia and industry~\cite{DBLP:journals/tsc/ChenB17,DBLP:journals/tse/ChenB17,DBLP:journals/tosem/ChenLBY18,DBLP:conf/icse/Chen19b,DBLP:conf/nips/KrizhevskySH12,DBLP:conf/issta/Liu0020}. Over the last decade, a variety of DL framework projects, such as TensorFlow and PyTorch, have been developed to enable rapid and seamless development of DL based software.

As with traditional software projects, DL frameworks inevitably contain bugs, especially those bugs that are related to performance (e.g., poor user experience, degraded responsiveness, and waste computational resources)~\cite{DBLP:conf/asplos/WenLBC18} and accuracy (e.g., insufficient prediction outcomes and loss)~\cite{DBLP:conf/kbse/FrancoGR17}. Indeed, these \bugtype bugs in DL frameworks can lead to severe consequences, affecting any software that is built on top of them~\cite{DBLP:conf/kbse/GuoXLZLLS20}. For example, according to the U.S. National Transportation Safety Board (NTSB), the recent accident of Uber's self-driving car was caused by \bugtype bugs of their DL framework, which inaccurately classified a pedestrian as an unknown object under specific conditions and doing so with a slow response\footnote{\url{https://tinyurl.com/ykufbpey}.}. Therefore, to improve the quality of continuous maintenance, mainstreamed DL frameworks make use of modern tracking systems --- most commonly GitHub --- to allow bugs to be reported, discussed, and eventually fixed. This paper focuses on understanding such a practice on the life-cycle of reporting \bugtype bugs for DL frameworks.

It has been well-recognized that the bug report analysis is at least as difficult as the actual bug-fixing~\cite{DBLP:conf/cascon/AntoniolAPKG08a,DBLP:conf/kbse/HanYL18,DBLP:journals/ese/TianLXS15}. In fact, \citeauthor{DBLP:conf/eclipse/AnvikHM05}~\cite{DBLP:conf/eclipse/AnvikHM05} discover that the developers and maintainers are often overwhelmed with a large number of bug reports. The task becomes even more time-consuming when it comes to understanding the content: \citeauthor{DBLP:conf/icse/HerzigJZ13}~\cite{DBLP:conf/icse/HerzigJZ13} report that it takes at least 90 working days of efforts (for two experienced developers) to merely classify around 7,000 reports --- this does not even include extracting useful information from them. The reason could be partial because a bug report is often written in a considerable length, e.g., up to 332 sentences per report in average~\cite{DBLP:conf/sigsoft/ManiCSD12}. Furthermore, a submitted bug report may not be associated with an actual bug, which could only be known after inspection~\cite{DBLP:conf/icse/HerzigJZ13}. The analysis is particularly difficult for the performance and accuracy related bug reports since they are often implicit. That is, there is no precise oracle to assess them, meaning that it is typically hard to understand how ``slow'' or how ``inaccurate'' the results are would be considered as a bug without thorough investigation. As a result, insights on the state-of-the-practice for reporting bugs/concerns are as important as understanding the characteristics of an actual bug itself.

Previous work exists on understanding the causes of bugs and how they are fixed for traditional projects~\cite{DBLP:conf/asplos/LiTWLZZ06,DBLP:journals/ese/TanLLWZZ14}, but there is a lack of studies that target explicitly \bugtype bugs.
From another perspective, \citet{DBLP:journals/tse/ZimmermannPBJSW10} analyze the practice of how bugs are reported in traditional software from their classic tracking systems such as JIRA. However, since DL frameworks hold different stacks, fixed patterns, and software engineering practices from the traditional software projects~\cite{DBLP:conf/icse/IslamPNR20,DBLP:conf/icse/AmershiBBDGKNN019,DBLP:conf/aaai/SrisakaokulWAAX18}, the conclusions drawn on traditional projects are not necessarily applicable to the DL ones. Further, most of the DL frameworks are open-sourced hosted on GitHub, whose tracking system is much more flexible, but highly unstructured compared with the classic ones. There exist studies that seek to investigate the characteristics of bugs for DL systems built on top of the DL frameworks~\cite{DBLP:conf/apsec/SunZLHYL17,DBLP:conf/issta/ZhangCCXZ18,DBLP:conf/sigsoft/IslamNPR19,DBLP:conf/icse/ZhangXZLLY20,DBLP:conf/issre/ThungWLJ12}. However, they do not target the level of DL frameworks and there is still a lack of understanding on their bug reporting practice, particularly related to \bugtype concerns, which is our focus in this work.

To close such a gap, this work presents an exploratory study of 10 popular open-sourced DL frameworks from GitHub. In particular, we collect and analyze 664 high-quality representative samples of the \bugtype bug reports from a total population of 22,522. As such, we seek to provide a comprehensive understanding of five research questions related to the \bugtype bug reports for DL frameworks. Specifically, our key contributions are:

\begin{itemize}
    \item Several empirical findings that provide better understandings of reporting \bugtype bugs for DL frameworks. In particular, we discuss answers and evidence for the following research questions (RQs):
    
    \begin{itemize}
      \item[---] \textbf{RQ1:} What is the most common reason for reporting?
      
      \textbf{Answer:} ``low speed'' is the most common reason for submitting performance related bug reports (from 27\% to 67\% among the frameworks); however, we see no consistent pattern for accuracy related ones.
      \item[---] \textbf{RQ2:} Which DL stage(s) is the most relevant in reporting?
      
      \textbf{Answer:} The \textit{training} stage is prevalent in performance and accuracy bug reports, ranging between 38\% to 77\% across the frameworks.
      \item[---] \textbf{RQ3:} Do reports provide sufficient information for investigations?
      
      \textbf{Answer:} Yes, as states like ``not related'' or ``not enough information'' are rare cases.
      \item[---] \textbf{RQ4:} Are most reports bug-related?
      
      \textbf{Answer:} No, the majority (from 69\% up
to 100\%) of the closed performance and accuracy bug reports
are either unclassified or unrelated to actual bugs.
      \item[---] \textbf{RQ5:} Do bug-related reports always lead to patch(es)?
      
      \textbf{Answer:} In fact, around 50\% of the performance and accuracy bug reports, which indeed reveal bugs, are not resolved by direct patches.
    \end{itemize}

    \item Actionable implications to researchers, maintainers, and report submitters for the DL frameworks.
    
    \item A labeled dataset --- a six months effort --- that enables rapid proof-of-concept for future studies on problems related to \bugtype bug reports for DL. The dataset, together with other analyzed data, has been made publicly accessible via our online repository:
    {\texttt{\textcolor{blue}{\url{https://zenodo.org/record/6371676}}}}.
\end{itemize}

The rest of this paper is organized as follows. In Section~\ref{sec:bg}, we introduces the background information about \bugtype bug reports on GitHub. Section \ref{sec:method} describes the research methodology of our empirical study. Section~\ref{sec:data} elaborates the detailed data preparation process, following by an articulation of the classification criteria in Section~\ref{sec:class}. Section~\ref{sec:result} presents results and findings. The actionable implications derived from our findings are presented in Section~\ref{sec:implication}. In Section~\ref{sec:threat} and~\ref{sec:related}, we discuss the threats to validity and related work, respectively. Finally, Section~\ref{sec:con} draws a conclusion for this paper.

%% file: sec/bg.tex
\section{Background}
\label{sec:bg}

In this section, we introduce the necessary preliminaries for our empirical study.

\subsection{\Bugtype Bug Reports for DL Frameworks}

Indeed, despite being used primarily for bug reporting~\citep{DBLP:conf/apsec/SunZLHYL17}, the issues tracking system on GitHub can serve various purposes or even as a forum of discussion. However, those issues are formatted in a way that is intrinsically similar to the reports~\cite{DBLP:conf/issre/BissyandeLJRKT13}\footnote{Hence, whenever we call reports, we mean the issues on GitHub.}, including a title/summary, descriptions, comments, and labels. In this work, we are interested in analyzing those issues that are close to the ``bug reports'' in more traditional platforms, e.g., the JIRA. Specifically, our focus is the \textbf{\bugtype bug reports} --- those submitted issues that report observations or concerns about the undesired phenomena on the performance and accuracy of the DL framework, which may reveal \bugtype bugs~\citep{DBLP:conf/sigsoft/IslamNPR19}. We will elaborate on the inclusion and exclusion criteria used to extract the \bugtype bug reports in Section~\ref{sec:data}.

It is worth noting that the \bugtype bug reports merely express \textbf{\bugtype concerns}, which do not necessarily correlate with actual \textbf{\bugtype bugs}. That is, it may well be possible that the reports turn out to be some false alarms (due to, e.g., incorrect usage of the API) or they report something that can be easily resolved by using certain workarounds, which do not require a patch to fix. Indeed, given the open nature of the issue tracking system, one can submit a \orbugtype bug report as long as there is a \orbugtype concern, regardless of how trivial it is. This, together with the fact that the bug report itself is complicated to be analyzed~\cite{DBLP:conf/cascon/AntoniolAPKG08a,DBLP:conf/kbse/HanYL18,DBLP:journals/ese/TianLXS15,DBLP:conf/eclipse/AnvikHM05,DBLP:conf/sigsoft/ManiCSD12,DBLP:conf/icse/HerzigJZ13}, is the key reason why understanding the nature and life-cycle of \bugtype concerns/bugs reporting for DL frameworks is crucial, which is our focus in this work. 

\subsection{Bug Reports Lifecycle on GitHub}

\begin{figure}[t!]
  \centering
\includegraphics[width=\columnwidth]{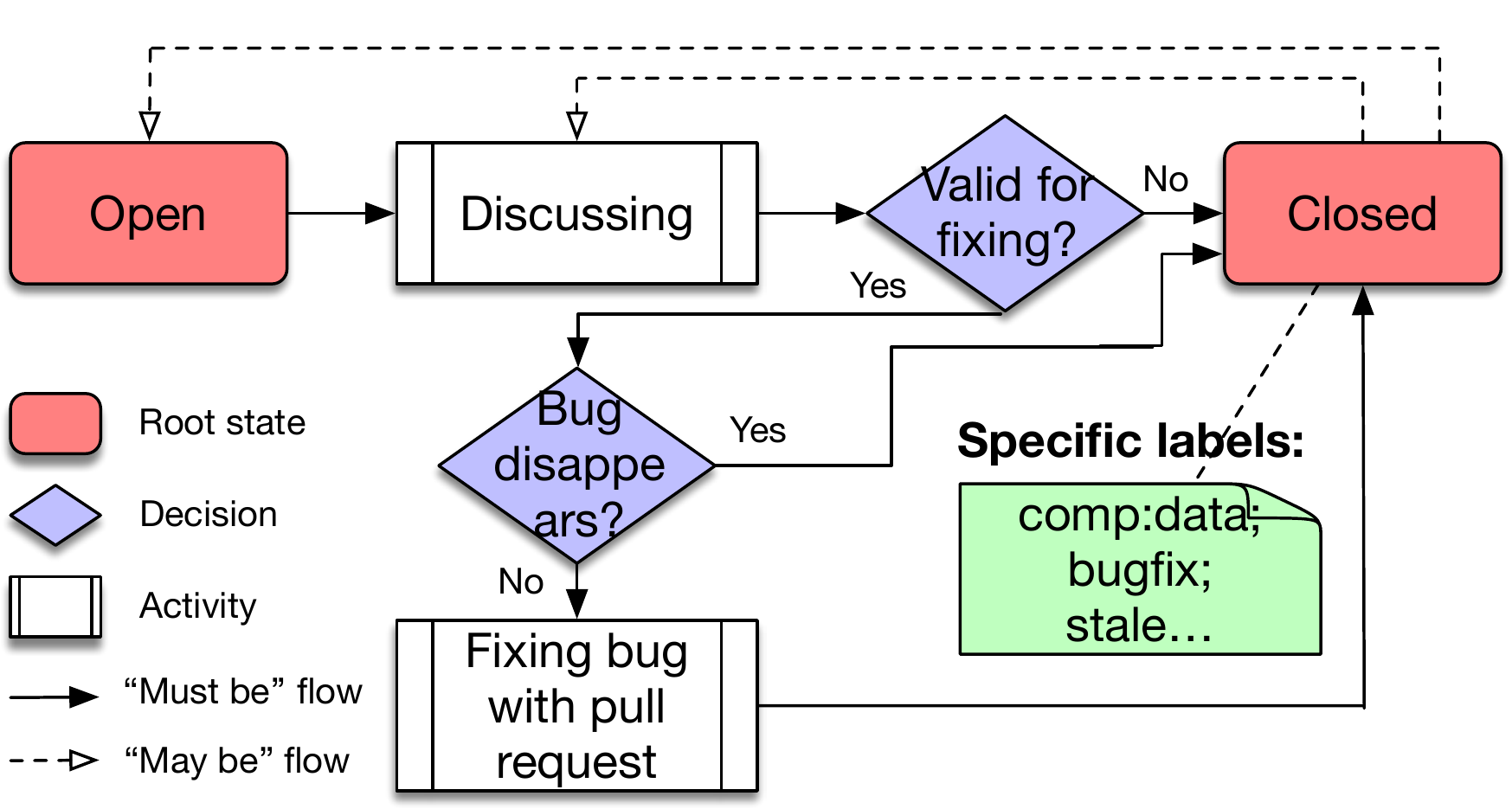}
 \caption{The brief lifecycle of bug reports on GitHub.}
 \label{fig:life}
\end{figure}

On GitHub, all the bug reports (regardless of whether they are \orbugtype specific) are committed to a standard lifecycle~\citep{DBLP:journals/chinaf/ZhangWHXZM15,DBLP:journals/air/UddinGDNS17}. As shown in Figure~\ref{fig:life}, a submitted bug report would undergo a discussion and commenting to identify whether it is valid for bug-fixing. If it does not involve a bug or cannot be determined, then the maintainers can close the report (with necessary conclusion and workarounds). If there is indeed a bug, a formal bug-fixing process would be triggered. During such a process, the bug may possibly disappear (e.g., being fixed unintentionally), in which case the report would be closed too. Otherwise, eventually, a directly associated patch(es) would be created with a pull request, awaiting approval of merge after which the report would be closed. Most commonly, a report is closed along with various custom labels added throughout the lifecycle of the bug report.

Note that the report may be reopened, as similar observations may occur again. Similarly, an extended discussion about the report is also possible even if the report has been closed.

%% file: sec/method.tex
\section{Methodology}
\label{sec:method}

\begin{figure}[t!]
  \centering
  \includegraphics[width=\columnwidth]{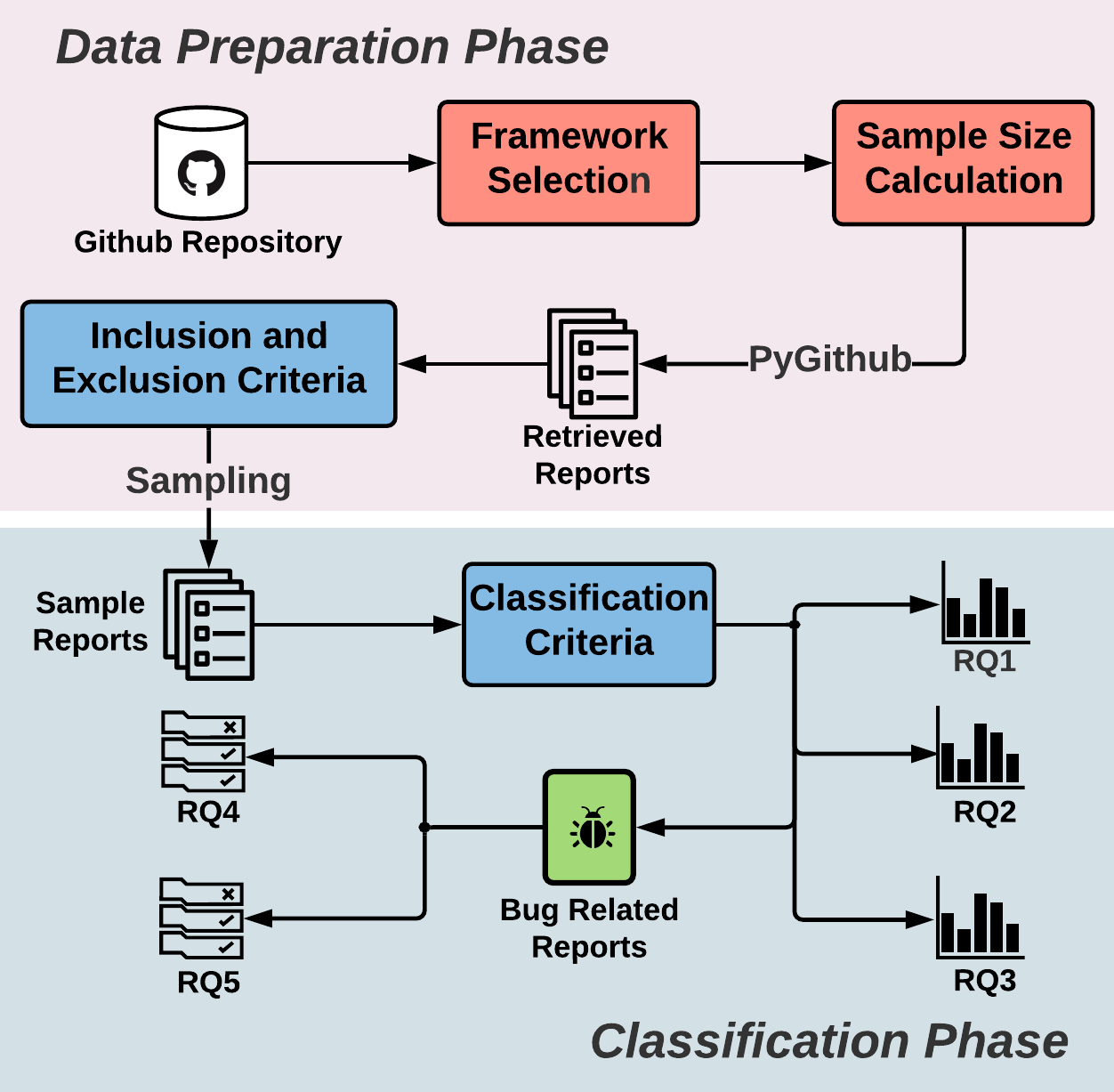}
  \caption{The empirical study methodology.}
  \label{fig:flow}
\end{figure}

As shown in Figure~\ref{fig:flow}, our research methodology consists of a \textit{Data Preparation Phase} and a \textit{Classification Phase}. 

During \textit{Data Preparation Phase}, we firstly conducted \textit{Framework Selection} to extract the most popular DL frameworks on GitHub according to the number of stars/folks. 

Next, we retrieve a total number of 22,522 reports (including those related to other types of bugs) for all frameworks over the period of five years using PyGithub Module\footnote{https://github.com/PyGithub/PyGithub}, for which is impractical to thoroughly analyze. Yet, according to the guidance from \citet{kadam2010sample}, we need at least 664 samples of \bugtype bug reports to gain meaningful interpretation at 99\% confidence level under such a population (see Section~\ref{sec:sample}). Therefore, we randomly sampled from the 22,522 reports until the collected number of \bugtype bug reports reached 664 according to the \textit{Inclusion and Exclusion Criteria}. For each framework, the sampled number is proportional to its percentage of the returned number of reports within the 22,522. This resulted in a manual inspection of more than 10,000 reports, including error-checking to ensure that none of the 664 \bugtype bug reports was misclassified, over the course of six months.

In the \textit{Classification Phase}, we classified the 664 sampled \bugtype bug reports according to the defined \textit{Classification Criteria}, which were derived from both the results of sampled reports and knowledge from prior work~\cite{DBLP:conf/issta/ZhangCCXZ18,DBLP:conf/sigsoft/IslamNPR19} (see Section~\ref{sec:class}). To avoid bias, we ensure that the Cohen's Kappa coefficient ($\kappa$)~\citep{mchugh2012interrater} of the classification is at least 0.7 between the authors (which means a substantial agreement~\citep{mchugh2012interrater}). The results lead to our answers to \textbf{RQ1}-\textbf{RQ3}, together with the reports that indeed reveal \bugtype bugs. From those bug-related reports, we can then draw findings for \textbf{RQ4} and \textbf{RQ5}.

%% file: sec/data.tex
\section{Data Preparation}
\label{sec:data}

We conducted the data collection in Jan 2021. Here, we specify the detailed steps of the data preparation process.

\subsection{Framework Selection}

To select the DL frameworks for this study, we mined the most popular ones from Github (based on the number of stars and folks). The only criterion we used is that the framework should not tie to a specific application domain of DL. As such, popular frameworks like Theano, OpenCV, and Torch7 were omitted as they focus specifically on Computer Vision. We eventually chose 10 most widely used DL frameworks, as shown in Table~\ref{tab:allocation}.

\subsection{Sampling Method and Size}
\label{sec:sample}
Using the PyGithub module for all 10 DL frameworks on GitHub, we retrieve a total number of 22,522 reports submitted between 1st Jan 2014 and 31st Dec 2019. \textcolor{black}{We chose this period as it contains a more balanced number of open and closed reports.} Since this is an extremely large number of reports, we wish to sample a set of high-quality representatives, ensuring that our conclusions would generalize to the whole population of each framework. To that end, we calculate the proper sample size following the guidance offered by \citet{kadam2010sample}:
\begin{equation}
{{N \times \varphi} \over {N + \varphi}}\text{, \hspace{0.3cm} subject to \hspace{0.1cm}} \varphi = {{z^{2} \times p \times (1-p)} \over {e^{2}}}
\end{equation}
where $N$ is the total number of reports (i.e., 22,522); $z$ is the two-sided z-score at a confidence level of 99\%; $e$ is the corresponding margin of error and $p$ is the proportion of \bugtype bug reports in the entire population (using the most conservative value 0.5). The above has led to 664 as the total sample amount for \bugtype bug reports. Then, we proportionally and randomly sample the \bugtype bug reports in different frameworks and states, according to the ratio between its total number for a framework/state and the total amount of searched reports for all frameworks, as shown in Table~\ref{tab:allocation}. This is important as there is an imbalanced distribution of the number of reports across the DL frameworks. Note that during the process, we ensure that the selection is completely random --- every report will have an equal chance to be selected. Again, all authors are involved in the process to improve reliability. In case of disagreement, the reports were investigated multiple times or counseling external experts until a consensus has been reached.

\subsection{Inclusion and Exclusion Criteria}

When sampling reports, we use the following inclusion criteria to decide whether a report should be considered as a \orbugtype bug report:

\begin{itemize}
    \item The report contains at least one clear symptom of performance or accuracy concern, such as the hang, unexpected loss, and slow speed (see next section).
    \item The report describes an observation, concern, or problem about the undesired phenomena on the performance or accuracy aspects of using the DL framework.
    \item The report has at least one label (in addition to open/closed).
\end{itemize}
We remove the report if it fits any of the exclusion criteria:

\begin{itemize}
    \item The report is related to documentation or a tutorial.
    \item It is a request for completely new features, despite being performance or accuracy-related.
    \item It is a thread of pure discussion, user feedback, or records of planned ``TODO'' tasks.
    \item It describes an error that causes a crash when using the DL framework.
\end{itemize}

\begin{table}[t!]
\centering

  \caption{Distribution of 664 samples for each DL project/state.}
  \label{tab:allocation}
  \setlength{\tabcolsep}{0.9mm}
  \begin{adjustbox}{max width=\columnwidth}
  \begin{tabular}{lccc||ccc}
    \toprule
    \textbf{Framework} & \textbf{\makecell{Retrieved\\Open}} & \textbf{\makecell{Retrieved\\Closed}} & \textbf{\makecell{Retrieved\\Total}} & \textbf{\makecell{Sampled\\Open}} & \textbf{\makecell{Sampled\\Closed}} & \textbf{\makecell{Sampled\\Total}} \\
    \midrule
    TensorFlow & 1657 & 7814 & 9471 & 49 & 230 & 279 \\
  \rowcolor{gray!20} Keras & 1392 & 3215 & 4607 & 41 & 94 & 135 \\
    PyTorch & 1123 & 2094 & 3217 & 33 & 62 & 95 \\
    \rowcolor{gray!20} MXNet & 445 & 1771 & 2216 & 13 & 52 & 65 \\
    Caffe & 178 & 957 & 1135 & 5 & 28 & 33 \\
    \rowcolor{gray!20} CNTK & 187 & 506 & 693 & 6 & 15 & 21 \\
    Chainer & 2 & 424 & 426 & 0 & 13 & 13 \\
    \rowcolor{gray!20} Darknet & 341 & 83 & 424 & 10 & 3 & 13 \\
    Caffe2 & 106 & 93 & 199 & 3 & 3 & 6 \\
    \rowcolor{gray!20} Tiny-dnn & 66 & 68 & 134 & 2 & 2 & 4 \\
    \textbf{Total} & 5497 & 17025 & 22522 & 162 & 502 & 664 \\
  \bottomrule
\end{tabular}
\end{adjustbox}
\end{table}

%% file: sec/class.tex
\section{Classification and Labeling}
\label{sec:class}

In this section, we present the criteria used to classify and label the sampled reports, which is the foundation of this study to derive and analyze our findings.

\subsection{Classification Criteria}
\subsubsection{Symptoms Claimed in DL \BugType Bug Reports}
\label{sec:class1}

To better study the reports, we summarize the following common symptoms of performance and accuracy for DL framework as inspired by the work of \citeauthor{DBLP:conf/issta/ZhangCCXZ18}~\cite{DBLP:conf/issta/ZhangCCXZ18}. In particular, the performance related symptoms are:

\begin{itemize}
    \item \textbf{Low Speed:} DL processing speed is rather slow in certain pipeline stages (e.g., training and prediction etc).
    \item \textbf{Abnormal Memory Usage:} The process consumes an abnormal amount of memory, e.g., too low, too high, or a possible leak. The case that leads to a crash is not included.
    \item \textbf{Hang:} The DL process is not responsive or runs indefinitely.
\end{itemize}

In contrast, the symptoms for accuracy are:

\begin{itemize}
    \item \textbf{Poor Loss:} Abnormal observation of the loss value during training, e.g., high loss, infinite loss, periodic loss, unchanged loss, and other unwanted loss values.
    \item \textbf{Poor Prediction:} Poor metric result during prediction (e.g., precision, recall, or accuracy) is observed.
    \item \textbf{Unexpected Output:} This occurs when the output of DL system contains, e.g., abnormal gradient value, abnormal weight value, unusual tensor calculation result.
\end{itemize}

\subsubsection{Stages of the DL Pipeline}

\label{sec:class2}
In this work, we classify the \bugtype bug reports into the following five key stages of typical DL systems when the frameworks are used to build them, as recommended by~\citet{DBLP:conf/sigsoft/IslamNPR19}:

\begin{itemize}
    \item \textbf{Data Processing:} This stage is responsible for data loading, preprocessing, input, and output filtering that enable more effective model training.
    \item \textbf{Model Construction:} This is related to the choice of model and hyperparameter tuning\footnote{Unlike~\citet{DBLP:conf/sigsoft/IslamNPR19}, we consider hyperparameter tuning as part of Model Construction as we found that they are often mentioned together in a \orbugtype bug report.}.
    \item \textbf{Training:} This involves training a neural network such that a loss function is minimized across data samples. It is the core of DL pipeline and can be strongly tied with hardware resources, e.g., GPU and CPU parallelism.
    \item \textbf{Evaluation:} This is concerned with validating and evaluating the quality of the model trained. Common metrics such as AUC, F-measure, or RMSE would be used here.
    \item \textbf{Prediction:} This is the stage where the DL system with a trained model is actually deployed in production, forecasting the outcome with newly given data.
\end{itemize}

\vspace{-0.1cm}
\vspace{-0.1cm}
\subsubsection{States of \BugType Bug Reports}
\label{sec:class3}
We found that most of the custom labels in GitHub are not states; even for those which indeed represent states, the majority of them are duplicate or represent similar meaning. Therefore, in what follow we summarize 11 states across the DL frameworks under which the report is closed:

\begin{itemize}
    \item \textbf{Fixed:} The report where a related patch is directly created or there is a claim that the bug is fixed in another release.
    
    \item \textbf{Resolved:} The state where an accepted workaround has been provided or the submitter discovers an alternative resolution. However, no change needs to be made to the codebase.
    \item \textbf{Not reproducible:} This means the reported observation has been found as difficult to be reproduced.
    \item \textbf{Not enough information:} This is often a closed report where the provided information has been claimed as too vague to generate discussion and investigation.
    \item \textbf{Not related:} The report has been confirmed to be unrelated to the DL framework.
    \item \textbf{No reason:} The submitter closes the report without giving any comment.
    \item \textbf{Better ask in elsewhere:} The maintainers suggest that the report is not suitable to be discussed on GitHub.
    \item \textbf{Working as expected:} The report is identified as not a concern, but merely about DL code design.
    \item \textbf{Lack of activity:} Closed by issue management system due to being idle over a period of time.
    \item \textbf{Stale:} The report has been identified by a maintainer as stale, hence should be closed.
    \item \textbf{Duplicate:} The report is closed as the same content has already been reported in an origin one.
 
\end{itemize}
In contrast, there is only one state to represent open reports:
\begin{itemize}
   \item \textbf{Open:} The \bugtype bug report has yet reached a conclusion about the next stage.
\end{itemize}
\vspace{-0.1cm}

\subsubsection{Correlation between Reports and Bugs}
\label{sec:class4}
Drawing on the states, we are able to easily summarize the \bugtype bug reports into the three categories below\footnote{\textcolor{black}{For the very small amount of reports with a \textit{duplicate} state, we use the state of its master bug report.}}: 
\begin{itemize}
    \item \textbf{Bug-related:} The reports under the state of \textit{fixed} are confirmed related to \bugtype bug by the maintainers, as it indicates that the reported observation has revealed a bug that triggers a bug-fixing process. 

    \item  \textbf{Bug-unrelated:} The reports are closed without associated fixes while having the states of \textit{resolved}, \textit{not reproducible}, \textit{not enough information}, \textit{not related}, \textit{better ask in elsewhere}, \textit{working as expected} are regarded as unrelated to \bugtype bugs by the maintainers, since they do not reveal any actual bugs.
    \item  \textbf{Unclassified:} \textit{No reason}, \textit{lack of activity}, and \textit{stale} states mean a report provides no information for maintainers to determine whether a bug is involved or not. 
\end{itemize}

For those bug-related reports, we further classify them depending on how the corresponding \bugtype bug is fixed: 

\begin{itemize}
\item \textbf{Fixed by patch(es):} This means that the bug is fixed by directly merging a patch(es) into the codebase.

\item \textbf{Fixed in newer release:} This refers to the report where there is no directly associated patch(es), but it was commented that the bug disappears in a newer release, implying that it must have been indirectly fixed as part of some other patches or refactoring. Note that this implies that the reported \bugtype concerns were fixed even without being raised by a report. 
\end{itemize}

All authors of this paper labeled the reports based on the classification criteria above by interpreting the content and comments of each report. Disagreements were resolved internally or by counseling external experts when needed. From these, we achieve a Cohen's Kappa coefficient $\kappa \in [0.7, 1.0]$ in each corresponding criteria after labeling, which indicates a substantial agreement~\citep{mchugh2012interrater}.

%% file: sec/result.tex
\section{Results}
\label{sec:result}

In this section, we present the results of our empirical study and answer the research questions posed in Section~\ref{sec:intro}. Note that to avoid bias, we analyze each DL framework individually and draw conclusions therein, since they have different total numbers of reports. The dataset and raw data are made available at: {\texttt{\textcolor{blue}{\url{https://zenodo.org/record/6371676}}}}.

\input{tab/rq1}

\subsection{Reasons of Reporting (RQ1)}

As can be seen from Table~\ref{tab:consequence} (two left-most columns), the concerns over performance and accuracy tend to be balanced across the DL frameworks. Looking into more detailed reasons of both performance or accuracy related bug reports, we see that for a majority of cases, the developers submit \bugtype bug reports mainly due to \textit{low speed} (8 out of 10, up to 67\%), especially for popular frameworks such as TensorFlow, MXNet, Chainer, and Caffe2. This is surprising, as despite the accuracy is a unique and key attribute for DL frameworks, the primary concern remains on the performance, i.e., time-related attributes.

Another observation is that, for the three concrete reasons under the performance concern, there is a strong bias towards the symptom of \textit{low speed}. In contrast, the concrete reasons of accuracy concern are relatively more balance and we cannot conclude which one is more prevalent.

Compared with the others, we found that to what extent can be considered as low speed in a performance bug report of DL frameworks is often more vaguely defined and with significantly different aspirations based on the context. For example, report \texttt{\#3996} for Keras reports that the ``\texttt{Model.fit takes about 10 minutes before it actually starts doing anything}'', upon which is considered unacceptable. In contrast, report \texttt{\#33340} for TensorFlow states that a prediction speed of 29ms is already a major slowdown. Such vagueness and diverse aspirations could explain why \textit{low speed} is the main reason for submitting a \orbugtype bug report.

Therefore we say:

\begin{quotebox}
\noindent
\textit{\textbf{Finding 1:} For DL frameworks, \textit{low speed} is the most prevalent concrete reason for submitting a performance related bug report (from 27\% up to 67\%). For accuracy related ones, we see no definitive patterns on the reason. }
\end{quotebox}

\input{tab/rq2}

\subsection{Reported Learning Stages (RQ2)}

From Table~\ref{tab:stage}, it is clear that all the key stages in DL are associated with the \bugtype bug reports. However, the majority of them are related to the \textit{training} stage of the DL pipeline for 8 out of 10 frameworks, all of which are the top most popular DL frameworks we found. We also note that the \textit{data processing} and \textit{model construction} tend to be two of the most uncommon stages in the \bugtype bug reports submitted.


We found that the \bugtype bug reports related to \textit{training} do not only predominate, but also lead to some of the most serious consequences. For example, report \texttt{\#9873} on PyTorch reports that ``\texttt{PyTorch is slow when only using CPU, and cannot utilize multicore of CPU}'', causing it to become about 30\% slower than Keras to train under the same condition.

In summary, we conclude that:

\begin{quotebox}
\textit{\textbf{Finding 2:} The \textit{training} stage is significantly more predominately concerned and relevant than the remaining four stages as stated in the \bugtype bug reports, constituting between 38\% and 77\%.}
\end{quotebox}

\subsection{Report States (RQ3)}

In Table~\ref{tab:stat}, most commonly a \orbugtype bug report is under an \textit{open} or \textit{resolved} state, as they are the most (or second most) prevalent for 7 frameworks. Note that a \textit{resolved} is different from a \textit{fixed}, as the former does not trigger a process of bug fixing; in most cases, the report is resolved because a workaround is provided; or there is a claim that the observation does not exist anymore. For example, report \texttt{\#9026} on MXNet reports that training speed is extremely slow with NVIDIA V100 GPU. Further investigation suggested a workaround of using \texttt{DataLodear} that can feed data asynchronously instead, which then ``resolves'' bug report but there is not a ``fix'', since no actual patch has been generated. 

In particular, PyTorch and MXNet exhibit good balance on \textit{open}, \textit{fixed}, and \textit{resolved} state with high parentage, implying a particularly healthy ``report-then-address'' cycle on \bugtype-related concerns. Darknet, Caffe2, and Tiny-dnn have a much higher share on \textit{open} than \textit{fixed} and \textit{resolved}, suggesting inactive maintenance, despite they are used in practice. Chainer, on the other extreme, has no open \orbugtype bug reports, suggesting it is either doing extremely well or simply attracts only a rather small proportion of users.

\input{tab/rq3}

Note that \textit{not related}, \textit{not enough information}, \textit{duplicate}, and \textit{not reproducible} are rare states, suggesting a good sign that the \bugtype bug reports submitted are of high quality.

Therefore, we say:

\begin{quotebox}
\noindent
\textit{\textbf{Finding 3:} The \bugtype bug reports under the state of ``\textit{open}'' and ``\textit{resolved}'' are significantly more prevalent than the others, suggesting a healthy maintenance cycle of \bugtype concerns in DL frameworks.}
\end{quotebox}
\begin{quotebox}
\textit{\textbf{Finding 4:} In contrast, the reports in the state of ``\textit{not related}'', ``\textit{not enough information}'', ``\textit{duplicate}'', and ``\textit{not reproducible}'' are much more rare, meaning that the \bugtype bug reports often provide sufficient information in DL frameworks.}
\end{quotebox}

\subsubsection{Reopened reports and followups on closed reports}

Two related and interesting questions to answer are what happens after a report is closed under whatever specific state, and how common for a closed report to be reopened? To this end, we look at whether a closed \orbugtype bug report can still have a sustainable discussion and whether (non-trivially) reopening reports are common\footnote{We count each report exactly once even if it has multiple reopens and followups on multiple closed states.}.

\begin{figure}[t!]
  \centering
\includegraphics[width=\columnwidth]{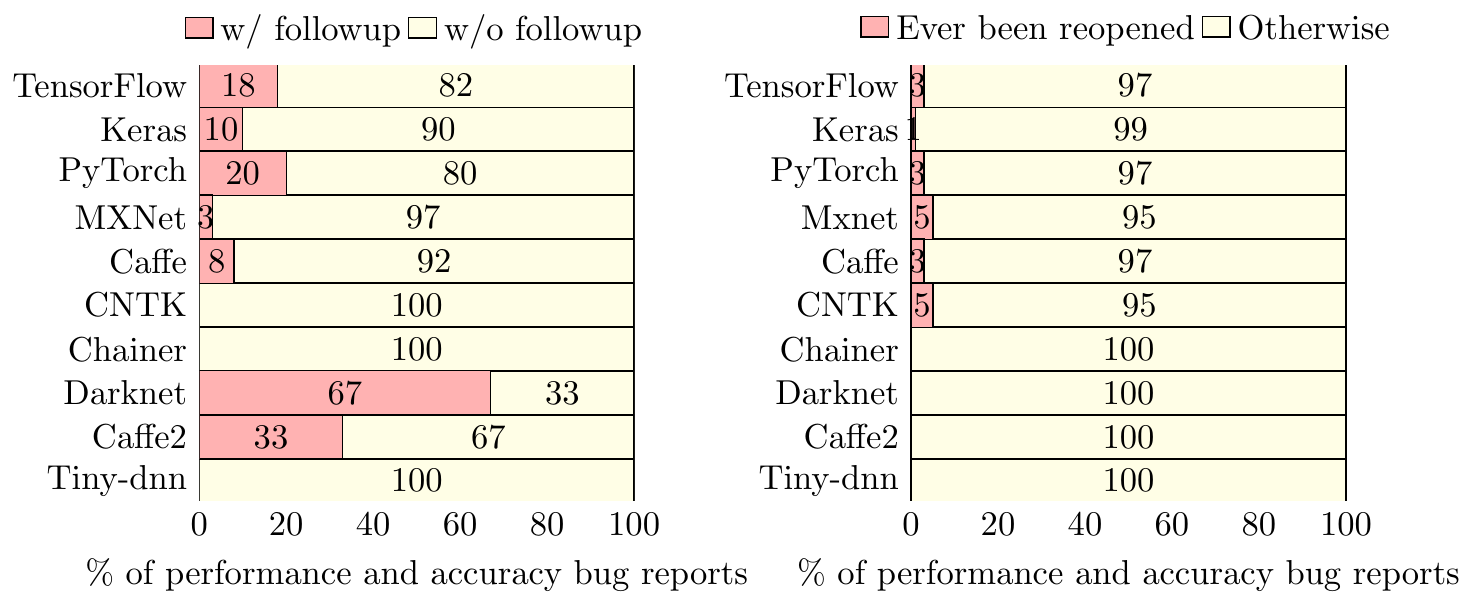}
  \caption{\% of closed \bugtype bug reports with/without followups (left) and whether they have ever been reopened (right).} 
 \label{fig:followup-reopen}
\end{figure}

From Figure~\ref{fig:followup-reopen} (left), we see that mostly the discussion of a \orbugtype bug report ends once the report is closed. However, there is also a good amount of them (up to 67\% on Darknet) where the new observations and discussion still continues without reopening them, since the proportions of the reports that have been ever reopened is significantly lower as shown in Figure~\ref{fig:followup-reopen} (right). Yet, despite rarely resulting in a reopening, such an extended discussion implies the importance/prevalence of the reported \bugtype observations/concerns, and then the report was closed without full satisfaction. In fact, the followups often lead to very positive outcomes. For example, in report \texttt{\#31243} for TensorFlow, it was reported that \texttt{tf.keras.load\_model} is very slow, the resolution proposed before closing the report is by installing \texttt{tf-nightly-gpu-2.0-preview}, which however poses some compatibility issues. In the extended dissuasion after the report was closed, a participant confirmed that the compatibility can cause a non-trivial issue, and a simpler workaround that loads the model from the function each time was suggested (by a different participant) and accepted.

Thus, we conclude that:

\begin{quotebox}
\noindent
\textit{\textbf{Finding 5:} For DL frameworks, it is not uncommon that follow-up observations/discussions are made to an already closed \orbugtype bug report. However, reopening a report is rare.}
\end{quotebox}

\subsection{Bug Revealing Reports (RQ4)}

For all the \bugtype bug reports that were closed in DL frameworks, Figure~\ref{fig:RQ4} shows how many of them can actually reveal at least one \orbugtype bug. Surprisingly, only small proportions of them are bug-related (between 11\% to 31\%). In contrast, it is most prevalent that a \orbugtype bug report is bug-unrelated (from 42\% up to 100\%), with those considered as unclassified ranked as the second most common. In particular, the number of unclassified reports is rare for TensorFlow, Keras, PyTorch, and Caffe2, while their proportions of bug-unrelated ones remain very high (with 100\% for Caffe2).

The above is a surprising sign that, albeit the \bugtype bug reports themselves in DL frameworks are generally of good quality, they do not often reveal actual \bugtype bugs that require bug-fixing.

From the above, we can summarize that:

\begin{quotebox}
\textit{\textbf{Finding 6:} In DL frameworks, the majority (from 69\% up to 100\%) of the closed \bugtype bug reports are either unrelated to actual \bugtype bugs (i.e., bug-unrelated) or unclassified.}
\end{quotebox}

\begin{figure}[t!]
  \centering
\includestandalone[width=\columnwidth]{fig/RQ4}
 \caption{\% of closed \bugtype bug reports that are bug-related, bug-unrelated, or unclassified.}
 \label{fig:RQ4}
\end{figure}

\begin{figure}[t!]
\centering
\includegraphics[width=\columnwidth]{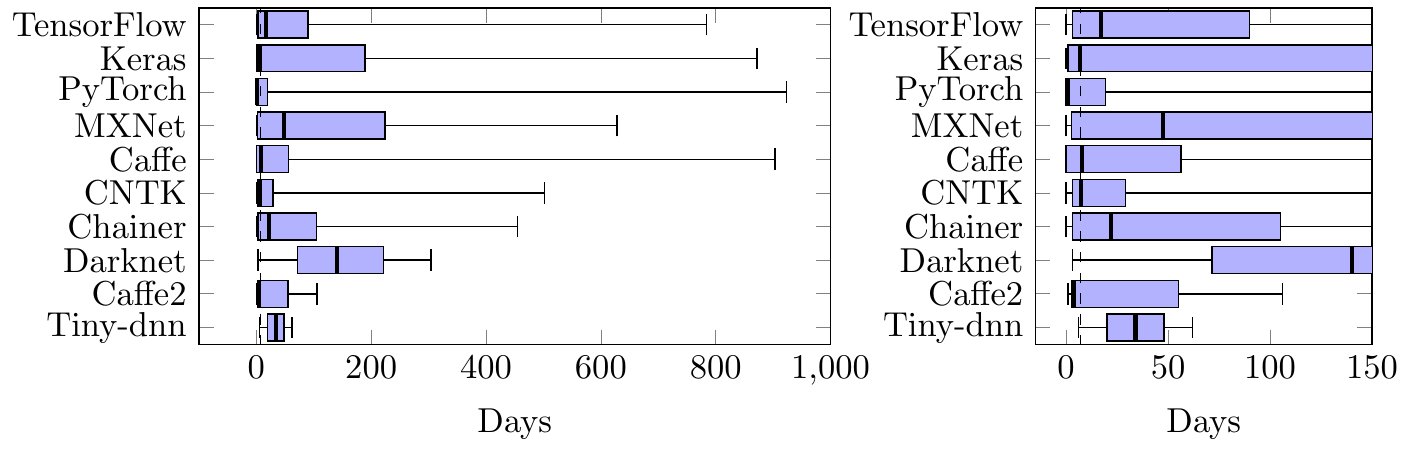}
\caption{Distribution of the time required (in days) for a \orbugtype bug report to be closed (dashed line denotes a week). The left shows an overall picture and the right is the ``zoom in'' version with a finer-grained scale.}
\label{fig:iden}
\end{figure}

\subsubsection{Time required}
A related question is how much time would be required to identify whether a \orbugtype bug report indeed reveals a bug? To this end, we further analyze, for all closed ones, the time (in days) taken between the reports being submitted and the comment that confirms its bug-relevance, i.e., whether it indeed discloses a \orbugtype bug. 

Figure~\ref{fig:iden} illustrates the results, in which we see that, for the median on 8 out of 10 frameworks, it needs a week or more for concluding whether the report is valid for bug-fixing. In particular, most of them exhibit a rather high interquartile range (e.g., Keras, MXNet, and Darknet), with the 75th percentile being more than 100 days. This suggests that the amount of time required (and potentially the efforts) is considerably high in general across the DL frameworks.

Therefore, we say:

\begin{quotebox}
\noindent
\textit{\textbf{Finding 7:} For DL frameworks, usually it takes at least a week to identify whether a \orbugtype bug report indeed reveals a bug.}
\end{quotebox}

\subsection{Patch(es) on Reports (RQ5)}

\begin{figure}[t!]
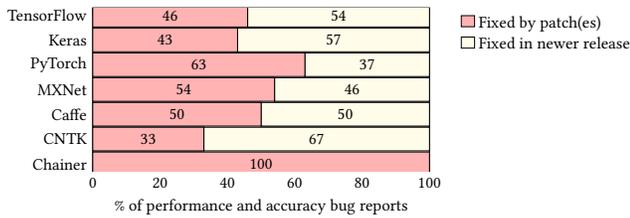

  \centering
\includestandalone[width=\columnwidth]{fig/RQ5-1}
  \caption{\% of fixed \bugtype bug reports that are fixed directly by patch(es) or indirectly in newer releases (Tiny-dnn, Caffe2, and Darknet are omitted as they have no fixed reports in our samples).} 
 \label{fig:RQ5-1}
\end{figure}

We seek to examine the distribution of whether a fix in a bug-related report is completed with a direct patch(es) or has already been dealt with in newer releases (as stated by the comments from the reports). As such, we consider only the reports under \textit{fixed} state. We do not consider those reports with a \textit{duplicate} state, as in those cases a \orbugtype bug is fixed when raised by another report. Therefore, when it is stated that a bug is fixed in a newer release, it often means that the bug was detected by the developer during routine refactoring rather than being raised from a report. As can be seen from Figure~\ref{fig:RQ5-1}, how a bug-related report is handled exhibits a reasonable balance between the two categories across all DL frameworks (except for Chainer). This is to our surprise, as it suggests that around half of the \bugtype bugs on DL frameworks were fixed without being formally raised in a report.

\begin{quotebox}
\noindent
\textit{\textbf{Finding 8:} Around half of the \bugtype bug reports in DL frameworks, which indeed reveal bugs, do not associate with a direct patch.}
\end{quotebox}

%% file: tab/rq1.tex
\begin{table}[t!]
\centering

\caption{\% on the reasons of submitting performance and accuracy bug reports for the DL frameworks (the most common one is highlighted). Note that a small amount of reports are linked with more than one reasons.}
  \label{tab:consequence}
\begin{adjustbox}{max width=\columnwidth}
\begin{tabular}{lcc||cccccc}
\toprule
\textbf{Frameworks} & \rotatebox{90}{\textbf{Performance}} & \rotatebox{90}{\textbf{Accuracy}} & \rotatebox{90}{\textbf{Low Speed}} & \rotatebox{90}{\textbf{Abnormal Memory Usage}} & \rotatebox{90}{\textbf{Hang}} & \rotatebox{90}{\textbf{Poor Loss}} & \rotatebox{90}{\textbf{Unexpected Output}} & \rotatebox{90}{\textbf{Poor Prediction}} \\
\midrule

Tensorflow & \cellcolor{blue!20}73\% & 27\% & \cellcolor{blue!20}52\% & 14\% & 6\% & 5\% & 13\% & 9\% \\
Keras & 38\% & \cellcolor{blue!20}64\% & \cellcolor{blue!20}32\% & 5\% & 1\% & 25\% & 8\% & 31\% \\
PyTorch & \cellcolor{blue!20}71\% & 29\% & \cellcolor{blue!20}44\% & 15\% & 12\% & 11\% & 17\% & 2\% \\
MXNet & \cellcolor{blue!20}62\% & 40\% & \cellcolor{blue!20}51\% & 8\% & 3\% & 5\% & 6\% & 29\% \\
Caffe & 39\% & \cellcolor{blue!20}61\% & \cellcolor{blue!20}27\% & 6\% & 6\% & 24\% & 21\% & 15\% \\
CNTK & \cellcolor{blue!20}64\% & 32\% & \cellcolor{blue!20}45\% & 18\% & 0\% & 5\% & 23\% & 5\% \\
Chainer & \cellcolor{blue!20}69\% & 31\% & \cellcolor{blue!20}62\% & 8\% & 0\% & 8\% & 15\% & 8\% \\
Darknet & 15\% & \cellcolor{blue!20}85\% & 15\% & 0\% & 0\% & \cellcolor{blue!20}38\% & 15\% & 31\% \\
Caffe2 & \cellcolor{blue!20}67\% & 33\% & \cellcolor{blue!20}67\% & 0\% & 0\% & 0\% & 33\% & 0\% \\
Tiny-dnn & 25\% & \cellcolor{blue!20}75\% & 25\% & 0\% & 0\% & 0\% & 25\% & \cellcolor{blue!20}50\%\\

\bottomrule
\end{tabular}
\end{adjustbox}
\end{table}

%% file: tab/rq2.tex
\begin{table}
\centering
  \caption{\% on the relevant DL stages to the submitted performance and accuracy bug reports (the most common one is highlighted). Note that a small amount of reports are linked with more than one stage.}
  \label{tab:stage}
\begin{adjustbox}{max width=\columnwidth}
\begin{tabular}{lcccccc}
\toprule
\textbf{Frameworks} & \textbf{\makecell{Data\\Processing}} & \textbf{\makecell{Model\\Construction}} & \textbf{Training} & \textbf{Evaluation} & \textbf{Prediction} \\
\midrule

Tensorflow & 18\% & 22\% & \cellcolor{blue!20}61\% & 13\% & 17\% \\
Keras & 5\% & 30\% & \cellcolor{blue!20}44\% & 16\% & 11\% \\
PyTorch & 33\% & 43\% & \cellcolor{blue!20}52\% & 37\% & 24\% \\
MXNet & 14\% & 18\% & \cellcolor{blue!20}65\% & 14\% & 14\% \\
Caffe & 3\% & 36\% & \cellcolor{blue!20}48\% & 6\% & 6\% \\
CNTK & 10\% & 24\% & \cellcolor{blue!20}38\% & 24\% & 5\% \\
Chainer & 15\% & 23\% & \cellcolor{blue!20}69\% & 8\% & 15\% \\
Darknet & 15\% & 0\% & \cellcolor{blue!20}77\% & 8\% & 0\% \\
Caffe2 & 0\% & 17\% & 33\% & 0\% & \cellcolor{blue!20}50\% \\
Tiny-dnn & 0\% & 0\% & 0\% & \cellcolor{blue!20}50\% & \cellcolor{blue!20}50\% \\

\bottomrule
\end{tabular}
\end{adjustbox}
\end{table}







%% file: tab/rq3.tex
\begin{table}[t!]
\centering
  \caption{\% on the mutually exclusive states of performance and accuracy bug reports in DL projects (the most common one is highlighted).}
  \label{tab:stat}
  \setlength{\tabcolsep}{1mm}
\begin{adjustbox}{max width=\columnwidth}
\begin{tabular}{lccccccccccccc}
\toprule

\textbf{Frameworks} & \rotatebox{90}{\textbf{Fixed}} & \rotatebox{90}{\textbf{Resolved}} & \rotatebox{90}{\textbf{Not reproducible}} & \rotatebox{90}{\textbf{Not enough information}} & \rotatebox{90}{\textbf{Not related}} & \rotatebox{90}{\textbf{No reason}} & \rotatebox{90}{\textbf{Better ask in elsewhere}} & \rotatebox{90}{\textbf{Working as expected}} & \rotatebox{90}{\textbf{Lack of activity}} & \rotatebox{90}{\textbf{Stale}} & \rotatebox{90}{\textbf{Duplicate}} & \rotatebox{90}{\textbf{Open}}\\

\midrule
Tensorflow & 12\% & \cellcolor{blue!20}29\% & 1\% & 2\% & 2\% & 10\% & 4\% & 5\% & 14\% & 1\% & 3\% & 18\% \\
Keras & 4\% & \cellcolor{blue!20}30\% & 0\% & 0\% & 1\% & 2\% & 0\% & 8\% & 2\% & 21\% & 1\% & \cellcolor{blue!20}30\% \\
PyTorch & 20\% & 19\% & 0\% & 0\% & 1\% & 5\% & 16\% & 1\% & 1\% & 1\% & 2\% & \cellcolor{blue!20}34\% \\
MXNet & 20\% & \cellcolor{blue!20}32\% & 0\% & 0\% & 0\% & 0\% & 2\% & 8\% & 17\% & 2\% & 0\% & 20\% \\
Caffe & 12\% & \cellcolor{blue!20}18\% & 6\% & 3\% & 3\% & 36\% & 0\% & 6\% & 0\% & 0\% & 0\% & 15\% \\
CNTK & 14\% & \cellcolor{blue!20}38\% & 0\% & 0\% & 0\% & 0\% & 0\% & 10\% & 5\% & 0\% & 0\% & 33\% \\
Chainer & \cellcolor{blue!20}31\% & 23\% & 0\% & 0\% & 8\% & 8\% & 0\% & 8\% & 0\% & 23\% & 0\% & 0\% \\
Darknet & 0\% & 15\% & 0\% & 0\% & 0\% & 0\% & 0\% & 8\% & 0\% & 0\% & 0\% & \cellcolor{blue!20}77\% \\
Caffe2 & 0\% & 33\% & 0\% & 0\% & 0\% & 0\% & 17\% & 0\% & 0\% & 0\% & 0\% & \cellcolor{blue!20}50\% \\
Tiny-dnn & 0\% & 25\% & 0\% & 0\% & 0\% & 0\% & 0\% & 0\% & 25\% & 0\% & 0\% & \cellcolor{blue!20}50\% \\
\bottomrule
\end{tabular}
\end{adjustbox}
\end{table}

%% file: fig/RQ4.tex
\begin{tikzpicture}
\begin{axis}[
 xbar stacked,
    width=8cm,
    height=6cm,
    xmax=100,
    xmin=0,
    bar width=0.45cm,
 enlarge y limits=0.05,
 enlarge x limits=0,
    	nodes near coords, 
  enlargelimits=0,
 enlarge y limits=0.05,
    legend style={at={(0.9,-0.07)},
      anchor=north,legend columns=1},
    xlabel={\% of performance and accuracy bug reports},
    legend style={draw=none,at={(1.25,1)}},
    legend cell align={left},
    symbolic y coords={J,I,H,G,F,E,D,C,B,A},
     yticklabels={TensorFlow, Keras, PyTorch, MXNet, Caffe, CNTK, Chainer, Darknet, Caffe2, Tiny-dnn},
         ytick=data
    ]

\addplot[xbar, fill=red!30, draw=black] coordinates {(21,A) (11,B) (31,C) (25,D) (14,E) (20,F) (31,G) (0,H) (0,I) (0,J)};

\addplot[xbar, fill=yellow!10, draw=black] coordinates {(75,A) (87,B) (65,C) (42,D) (79,E) (53,F) (38,G) (67,H) (100,I) (50,J)};

\addplot[xbar, fill=blue!50, draw=black] coordinates {(4,A) (2,B) (4,C) (33,D) (7,E) (27,F) (31,G) (33,H) (0,I) (50,J)};

\legend{Bug-related,Bug-unrelated,Unclassified}

\end{axis}
\end{tikzpicture}

%% file: fig/RQ5-1.tex
\begin{tikzpicture}
\begin{axis}[
 xbar stacked,
    width=8cm,
    height=4.7cm,
    xmax=100,
    xmin=0,
    bar width=0.45cm,
 enlarge y limits=0.05,
 enlarge x limits=0,
    	nodes near coords, 
  enlargelimits=0,
 enlarge y limits=0.05,
    legend style={at={(1.3,-0.07)},
      anchor=north,legend columns=1},
    xlabel={\% of performance and accuracy bug reports},
    legend style={draw=none,at={(1.35,1)}},
    legend cell align={left},
    symbolic y coords={J,I,H,G,F,E,D,C,B,A},
     yticklabels={TensorFlow, Keras, PyTorch, MXNet, Caffe, CNTK, Chainer},
         ytick=data
    ]

\addplot[xbar, fill=red!30, draw=black] coordinates {(46,A) (43,B) (63,C) (54,D) (50,E) (33,F) (100,G)};

\addplot[xbar, fill=yellow!10, draw=black] coordinates {(54,A) (57,B) (37,C) (46,D) (50,E) (67,F) (0,G)};

\legend{Fixed by patch(es), Fixed in newer release}

\end{axis}
\end{tikzpicture}

%% file: sec/implication.tex
\section{Actionable Implications}
\label{sec:implication}

We now discuss what actionable implications our empirical study and findings can provide to the practitioners of bug report analysis for DL frameworks.

\subsection{To Researchers}
As a first step, our empirical study provides clear motivations and the necessary data for researchers to investigate a wide range of related research problems on analyzing \bugtype bug reports for DL frameworks. In particular, our labeled dataset serves as the readily available foundation to efficiently provide proof-of-concept. Specifically, we provide the following implications:

\begin{enumerate}

     \item \textbf{Finding 5} reveals that discussion on a \orbugtype concern continues even after the report has been closed. However, little has been done to properly reflect the value and result of such discussion on the state of a report. It calls for future research to consider formalizing a more systematic protocol, or automated tools, that helps to make the decision on closing and reopening a \orbugtype bug report for DL frameworks.

    \item \textbf{Finding 6} shows that only a small proportion of the performance and accuracy bug reports are revealing \bugtype bugs. In the meantime, from \textbf{Finding 7}, we note that it often takes a considerably long time to identify whether a \orbugtype bug report can indeed reveal bugs. It is, therefore, desirable to have an automatic predictor that can identify which bug reports are worth bug-fixing, thus saving a significant amount of maintenance efforts. 

\end{enumerate}

\subsection{To Maintainers}

Deriving on the findings, we can provide the following actionable recommendations to the maintainers of the DL frameworks:

\begin{enumerate}
    \item \textbf{Finding 1} suggests that greater effort is still required to maintain, improve, or more correctly guide users to achieve the best training and prediction speed for DL frameworks.

    \item Since \textbf{Finding 2} reveals that \textit{training} is the more concerned and relevant DL stage in a \orbugtype bug report, hence the documentation or code related to \textit{training} requires more attention to maintain for \bugtype reasons. The goal is to reduce the change of bug reports in the first place, regardless of whether they are indeed revealing bugs or merely false alarms.

\end{enumerate}

\subsection{To Report Submitters}

Our findings also draw actionable suggestions to the report submitters for the DL frameworks:

\begin{enumerate}

\item \textbf{Finding 3} and \textbf{Finding 4} suggest that the current practice of writing \bugtype bug reports is healthy, therefore our results confirm no demand of any significant change.

\item \textbf{Finding 8} reveals that, for DL frameworks, around half of the \bugtype bug reports that can indeed reveal bugs do not lead to direct patches. Therefore, we suggest that before submitting a performance or accuracy related report, one should also examine the version at the newest ``release candidate'' branch, even if that is not a stable version. This would likely help reduce the bug-related reports that do not lead to an actual patch.

\end{enumerate}

%% file: sec/threat.tex
\section{Threats to Validity}
\label{sec:threat}

Threats to \textbf{internal validity} can be related to the classification of the \bugtype bug reports. We mitigate this in two steps: firstly, we codify the classification criteria based on either what has been well-acknowledged~\citep{DBLP:conf/issta/ZhangCCXZ18} or deriving from our samples, which involves all authors in order to reach common agreements. The identification of whether a report is a \orbugtype bug report also follows systematic inclusion and exclusion criteria, which is only confirmed once agreed by all authors. Secondly, when labeling the \bugtype bug reports, the inter-rater agreements were measured using Kappa coefficient ($\kappa$). From this, we achieve $\kappa \geq 0.7$ in all cases. However, we admit that errors may be inevitable during the manual process.

To ensure \textbf{construct validity} and avoid \textbf{conclusion bias} to particular DL frameworks, we consider the total number of reports for each (and those with the open and closed root state), we then conduct a random sample proportionally according to their totals. This helps us to better ensure a fair comparison under the imbalance distributions between frameworks and root states. In particular, upon reporting the results, we leverage the \% of individual frameworks whenever required, which further prevents the conclusions from being dominated by certain DL frameworks. 

The other threats can be related to the \textbf{external validity}, which is about the trustworthiness and generalizability of the conclusion drawn from the samples. We tackle this by investigating 10 DL frameworks with diverse characteristics and scales. For the actual sampling, we follow what has been recommended by \citet{kadam2010sample} to calculate the required sample size, which ensures that we are 99\% confident that the samples are representative enough for the whole population. Indeed, examining more DL frameworks and samples may provide more insights, but this is a rather expensive process as it has already taken around six months to analyze what we have collected in this paper.

%% file: sec/related.tex
\section{Related Work}
\label{sec:related}

We now discuss the prior work in light of the purpose and findings of our empirical study.

\textit{\textbf{Studies on Bugs for Projects based on Learning Frameworks:}} Since the modern era of Artificial Intelligence, there has been a few studies focusing on the characteristics of real bugs on projects based on learning frameworks~\citep{DBLP:conf/apsec/SunZLHYL17,DBLP:conf/issta/ZhangCCXZ18,DBLP:conf/sigsoft/IslamNPR19,DBLP:conf/icse/ZhangXZLLY20,DBLP:conf/issre/ThungWLJ12}. Among them, \citet{DBLP:conf/issre/ThungWLJ12} and \citet{DBLP:conf/icse/ZhangXZLLY20} focus on traditional bug report repositories, while \citet{DBLP:conf/apsec/SunZLHYL17}, \citet{DBLP:conf/sigsoft/IslamNPR19}, \citet{DBLP:conf/icse/HumbatovaJBR0T20} and \citet{DBLP:conf/issta/ZhangCCXZ18} work on data from GitHub commits/issues. Nevertheless, our work differs from the above on that:
\begin{itemize}
    \item Instead of analyzing the characteristics of the real bugs (as in the above work), we focus on understanding the practice of reporting \bugtype bugs, i.e., the \bugtype bug report itself, from GitHub. 
    \item Our purpose is to study the reports about \bugtype bugs as opposed to the general bugs (where most commonly the functional ones are the majority) from prior work. This allows us to draw more specific conclusions.
     \item We target the level of DL framework as opposed to the software that is built on top of it.
    \item We collect and analyze a moderate size of 664 samples from 10 DL frameworks. 
\end{itemize}

\textit{\textbf{Studies on Performance and Accuracy Bugs:}} While accuracy bugs are rarely studied for traditional software projects, performance bugs have been shown to be more critical and difficult to deal with compared with their functional counterparts, c.f.~\citep{DBLP:conf/esem/HanY16,DBLP:conf/msr/ZamanAH11}. For general open-sourced projects, a number of studies have been conducted to understand the cause, severity, and possible fix of real performance bugs~\citep{DBLP:conf/issta/Tizpaz-NiariC020,DBLP:conf/icse/Zhao0W0L19,DBLP:conf/issre/Chen0S19,DBLP:conf/wosp/Zhao0WS0LB20}, ranging from 109 to 700 samples, from classic bug repositories like JIRA. 

Studies of performance bugs on a specific domain of projects also exist. For example, \citet{DBLP:conf/icse/LiuXC14} investigate 70 performance bugs collected from eight Android projects. The results cover properties such as impacts, bug manifestation, debugging, and bug-fixing effort. Likewise, \citet{DBLP:conf/se/SelakovicP17} analyze 98 performance bugs from 16 client-side and server-side JavaScript projects. 
However, we focus on different purposes from those studies:

\begin{itemize}
    \item We investigate not only performance bug reports but also accuracy bug reports, which are rarely studied.
    \item Our empirical study focuses on the nature of \bugtype bug reports rather than the bugs themselves.
    \item We focus on DL frameworks, which have been shown that share little similarity to traditional projects~\citep{DBLP:conf/icse/AmershiBBDGKNN019,DBLP:conf/aaai/SrisakaokulWAAX18,DBLP:conf/icse/IslamPNR20}.
    \item We rely on GitHub that allows bugs to be reported without complying with more restricted rules compared with, e.g., JIRA. This imposes more difficulty in analysis.
\end{itemize}


\textit{\textbf{Studies on Bug Reports:}} Empirical studies focusing on the bug report itself have also been an important thread of research. With the target of traditional software projects, \citet{DBLP:journals/tse/ZimmermannPBJSW10} investigate what criteria can form a high-quality bug report that is most useful for bug-fixing. \citet{DBLP:conf/csmr/XiaLWSZ14} also empirically study how the bug reports are labeled, assigned, and given states. \citet{DBLP:conf/esem/ZhaoZSZH16} seek to understand whether there is a correlation between the discussion on a bug report and the quality of bug-fixing. Finally, by studying the Android-based projects, \citet{DBLP:conf/csmr/BhattacharyaUNK13} focus on how the quality of bug reports on different types of bugs can impact the developers' behavior. However, the above work did not target \bugtype bug reports for DL frameworks.

%% file: sec/con.tex
\vspace{-0.2cm}
\section{Conclusion and Future Work}
\label{sec:con}
In this work, we perform an empirical study that seeks to better understand the practice of reporting \bugtype bugs for DL frameworks. Our study systematically samples and analyzes 664 \bugtype bug reports from 22,522 issues over 10 DL frameworks on GitHub. The key findings are:

\begin{itemize}

\item ``low speed'' is the key reason for submitting performance related bug reports while the reason for reporting accuracy related concerns varies.

\item \textit{training} is the most prevalent DL stage in the \bugtype bug report.

\item the \bugtype bug reports are often of sufficient information.

\item majority of the \bugtype bug reports do not reveal actual bugs.

\item around half of the \bugtype bug reports, which reveal bugs, do not associate with the direct patches.

\end{itemize}

Drawing on the findings, we provide actionable implications to researchers, maintainers, and submitters involved in the \bugtype bug reporting process for DL frameworks.

With this paper, we hope to raise the importance of understanding the reporting practice of performance and accuracy bugs for DL frameworks. Indeed, by leveraging the dataset from this work, the aforementioned implications have also hinted at the necessity of possible future research threads, such as automatic tools on better \bugtype bug report identification.

%% file: main.bbl

\begin{thebibliography}{44}


\ifx \showCODEN    \undefined \def \showCODEN     #1{\unskip}     \fi
\ifx \showDOI      \undefined \def \showDOI       #1{#1}\fi
\ifx \showISBNx    \undefined \def \showISBNx     #1{\unskip}     \fi
\ifx \showISBNxiii \undefined \def \showISBNxiii  #1{\unskip}     \fi
\ifx \showISSN     \undefined \def \showISSN      #1{\unskip}     \fi
\ifx \showLCCN     \undefined \def \showLCCN      #1{\unskip}     \fi
\ifx \shownote     \undefined \def \shownote      #1{#1}          \fi
\ifx \showarticletitle \undefined \def \showarticletitle #1{#1}   \fi
\ifx \showURL      \undefined \def \showURL       {\relax}        \fi
\providecommand\bibfield[2]{#2}
\providecommand\bibinfo[2]{#2}
\providecommand\natexlab[1]{#1}
\providecommand\showeprint[2][]{arXiv:#2}

\bibitem[\protect\citeauthoryear{Amershi, Begel, Bird, DeLine, Gall, Kamar,
  Nagappan, Nushi, and Zimmermann}{Amershi et~al\mbox{.}}{2019}]%
        {DBLP:conf/icse/AmershiBBDGKNN019}
\bibfield{author}{\bibinfo{person}{Saleema Amershi}, \bibinfo{person}{Andrew
  Begel}, \bibinfo{person}{Christian Bird}, \bibinfo{person}{Robert DeLine},
  \bibinfo{person}{Harald~C. Gall}, \bibinfo{person}{Ece Kamar},
  \bibinfo{person}{Nachiappan Nagappan}, \bibinfo{person}{Besmira Nushi}, {and}
  \bibinfo{person}{Thomas Zimmermann}.} \bibinfo{year}{2019}\natexlab{}.
\newblock \showarticletitle{Software engineering for machine learning: a case
  study}. In \bibinfo{booktitle}{\emph{Proc. 41st International Conference on
  Software Engineering: Software Engineering in Practice}}.
\newblock


\bibitem[\protect\citeauthoryear{Antoniol, Ayari, Penta, Khomh, and
  Gu{\'{e}}h{\'{e}}neuc}{Antoniol et~al\mbox{.}}{2018}]%
        {DBLP:conf/cascon/AntoniolAPKG08a}
\bibfield{author}{\bibinfo{person}{Giuliano Antoniol}, \bibinfo{person}{Kamel
  Ayari}, \bibinfo{person}{Massimiliano~Di Penta}, \bibinfo{person}{Foutse
  Khomh}, {and} \bibinfo{person}{Yann{-}Ga{\"{e}}l Gu{\'{e}}h{\'{e}}neuc}.}
  \bibinfo{year}{2018}\natexlab{}.
\newblock \showarticletitle{Is it a bug or an enhancement?: a text-based
  approach to classify change requests}. In \bibinfo{booktitle}{\emph{Proc.
  28th Annual International Conference on Computer Science and Software
  Engineering, {CASCON} 2018}}. \bibinfo{publisher}{{ACM}}.
\newblock


\bibitem[\protect\citeauthoryear{Anvik, Hiew, and Murphy}{Anvik
  et~al\mbox{.}}{2005}]%
        {DBLP:conf/eclipse/AnvikHM05}
\bibfield{author}{\bibinfo{person}{John Anvik}, \bibinfo{person}{Lyndon Hiew},
  {and} \bibinfo{person}{Gail~C. Murphy}.} \bibinfo{year}{2005}\natexlab{}.
\newblock \showarticletitle{Coping with an open bug repository}. In
  \bibinfo{booktitle}{\emph{Proc. 2005 {OOPSLA} workshop on Eclipse Technology
  eXchange}}.
\newblock


\bibitem[\protect\citeauthoryear{Bhattacharya, Ulanova, Neamtiu, and
  Koduru}{Bhattacharya et~al\mbox{.}}{2013}]%
        {DBLP:conf/csmr/BhattacharyaUNK13}
\bibfield{author}{\bibinfo{person}{Pamela Bhattacharya},
  \bibinfo{person}{Liudmila Ulanova}, \bibinfo{person}{Iulian Neamtiu}, {and}
  \bibinfo{person}{Sai~Charan Koduru}.} \bibinfo{year}{2013}\natexlab{}.
\newblock \showarticletitle{An Empirical Analysis of Bug Reports and Bug Fixing
  in Open Source Android Apps}. In \bibinfo{booktitle}{\emph{17th Conference on
  Software Maintenance and Reengineering}}.
\newblock


\bibitem[\protect\citeauthoryear{Bissyand{\'{e}}, Lo, Jiang,
  R{\'{e}}veill{\`{e}}re, Klein, and Traon}{Bissyand{\'{e}}
  et~al\mbox{.}}{2013}]%
        {DBLP:conf/issre/BissyandeLJRKT13}
\bibfield{author}{\bibinfo{person}{Tegawend{\'{e}}~F. Bissyand{\'{e}}},
  \bibinfo{person}{David Lo}, \bibinfo{person}{Lingxiao Jiang},
  \bibinfo{person}{Laurent R{\'{e}}veill{\`{e}}re}, \bibinfo{person}{Jacques
  Klein}, {and} \bibinfo{person}{Yves~Le Traon}.}
  \bibinfo{year}{2013}\natexlab{}.
\newblock \showarticletitle{Got issues? Who cares about it? {A} large scale
  investigation of issue trackers from GitHub}. In
  \bibinfo{booktitle}{\emph{{IEEE} 24th International Symposium on Software
  Reliability Engineering, {ISSRE} 2013}}.
\newblock


\bibitem[\protect\citeauthoryear{Chen}{Chen}{2019}]%
        {DBLP:conf/icse/Chen19b}
\bibfield{author}{\bibinfo{person}{Tao Chen}.} \bibinfo{year}{2019}\natexlab{}.
\newblock \showarticletitle{All versus one: an empirical comparison on
  retrained and incremental machine learning for modeling performance of
  adaptable software}. In \bibinfo{booktitle}{\emph{Proceedings of the 14th
  International Symposium on Software Engineering for Adaptive and
  Self-Managing Systems}}. \bibinfo{pages}{157--168}.
\newblock


\bibitem[\protect\citeauthoryear{Chen and Bahsoon}{Chen and Bahsoon}{2017a}]%
        {DBLP:journals/tse/ChenB17}
\bibfield{author}{\bibinfo{person}{Tao Chen} {and} \bibinfo{person}{Rami
  Bahsoon}.} \bibinfo{year}{2017}\natexlab{a}.
\newblock \showarticletitle{Self-Adaptive and Online QoS Modeling for
  Cloud-Based Software Services}.
\newblock \bibinfo{journal}{\emph{{IEEE} Trans. Software Eng.}}
  \bibinfo{volume}{43}, \bibinfo{number}{5} (\bibinfo{year}{2017}),
  \bibinfo{pages}{453--475}.
\newblock


\bibitem[\protect\citeauthoryear{Chen and Bahsoon}{Chen and Bahsoon}{2017b}]%
        {DBLP:journals/tsc/ChenB17}
\bibfield{author}{\bibinfo{person}{Tao Chen} {and} \bibinfo{person}{Rami
  Bahsoon}.} \bibinfo{year}{2017}\natexlab{b}.
\newblock \showarticletitle{Self-Adaptive Trade-off Decision Making for
  Autoscaling Cloud-Based Services}.
\newblock \bibinfo{journal}{\emph{{IEEE} Trans. Serv. Comput.}}
  \bibinfo{volume}{10}, \bibinfo{number}{4} (\bibinfo{year}{2017}),
  \bibinfo{pages}{618--632}.
\newblock


\bibitem[\protect\citeauthoryear{Chen, Li, Bahsoon, and Yao}{Chen
  et~al\mbox{.}}{2018}]%
        {DBLP:journals/tosem/ChenLBY18}
\bibfield{author}{\bibinfo{person}{Tao Chen}, \bibinfo{person}{Ke Li},
  \bibinfo{person}{Rami Bahsoon}, {and} \bibinfo{person}{Xin Yao}.}
  \bibinfo{year}{2018}\natexlab{}.
\newblock \showarticletitle{{FEMOSAA:} Feature-Guided and Knee-Driven
  Multi-Objective Optimization for Self-Adaptive Software}.
\newblock \bibinfo{journal}{\emph{{ACM} Trans. Softw. Eng. Methodol.}}
  \bibinfo{volume}{27}, \bibinfo{number}{2} (\bibinfo{year}{2018}),
  \bibinfo{pages}{5:1--5:50}.
\newblock


\bibitem[\protect\citeauthoryear{Chen, Winter, and Suri}{Chen
  et~al\mbox{.}}{2019}]%
        {DBLP:conf/issre/Chen0S19}
\bibfield{author}{\bibinfo{person}{Yiqun Chen}, \bibinfo{person}{Stefan
  Winter}, {and} \bibinfo{person}{Neeraj Suri}.}
  \bibinfo{year}{2019}\natexlab{}.
\newblock \showarticletitle{Inferring Performance Bug Patterns from Developer
  Commits}. In \bibinfo{booktitle}{\emph{30th {IEEE} International Symposium on
  Software Reliability Engineering, {ISSRE} 2019}}.
\newblock


\bibitem[\protect\citeauthoryear{Franco, Guo, and
  Rubio{-}Gonz{\'{a}}lez}{Franco et~al\mbox{.}}{[n.d.]}]%
        {DBLP:conf/kbse/FrancoGR17}
\bibfield{author}{\bibinfo{person}{Anthony~Di Franco}, \bibinfo{person}{Hui
  Guo}, {and} \bibinfo{person}{Cindy Rubio{-}Gonz{\'{a}}lez}.}
  \bibinfo{year}{[n.d.]}\natexlab{}.
\newblock \showarticletitle{A comprehensive study of real-world numerical bug
  characteristics}. In \bibinfo{booktitle}{\emph{Proceedings of the 32nd
  {IEEE/ACM} International Conference on Automated Software Engineering}}.
  \bibinfo{pages}{509--519}.
\newblock


\bibitem[\protect\citeauthoryear{Goodfellow, Bengio, Courville, and
  Bengio}{Goodfellow et~al\mbox{.}}{2016}]%
        {goodfellow2016deep}
\bibfield{author}{\bibinfo{person}{Ian Goodfellow}, \bibinfo{person}{Yoshua
  Bengio}, \bibinfo{person}{Aaron Courville}, {and} \bibinfo{person}{Yoshua
  Bengio}.} \bibinfo{year}{2016}\natexlab{}.
\newblock \bibinfo{booktitle}{\emph{Deep learning}}. Vol.~\bibinfo{volume}{1}.
\newblock \bibinfo{publisher}{MIT press Cambridge}.
\newblock


\bibitem[\protect\citeauthoryear{Guo, Xie, Li, Zhang, Liu, Li, and Shen}{Guo
  et~al\mbox{.}}{2020}]%
        {DBLP:conf/kbse/GuoXLZLLS20}
\bibfield{author}{\bibinfo{person}{Qianyu Guo}, \bibinfo{person}{Xiaofei Xie},
  \bibinfo{person}{Yi Li}, \bibinfo{person}{Xiaoyu Zhang},
  \bibinfo{person}{Yang Liu}, \bibinfo{person}{Xiaohong Li}, {and}
  \bibinfo{person}{Chao Shen}.} \bibinfo{year}{2020}\natexlab{}.
\newblock \showarticletitle{Audee: Automated Testing for Deep Learning
  Frameworks}. In \bibinfo{booktitle}{\emph{35th {IEEE/ACM} International
  Conference on Automated Software Engineering, {ASE} 2020, Melbourne,
  Australia, September 21-25, 2020}}. \bibinfo{publisher}{{IEEE}},
  \bibinfo{pages}{486--498}.
\newblock


\bibitem[\protect\citeauthoryear{Han and Yu}{Han and Yu}{2016}]%
        {DBLP:conf/esem/HanY16}
\bibfield{author}{\bibinfo{person}{Xue Han} {and} \bibinfo{person}{Tingting
  Yu}.} \bibinfo{year}{2016}\natexlab{}.
\newblock \showarticletitle{An Empirical Study on Performance Bugs for Highly
  Configurable Software Systems}. In \bibinfo{booktitle}{\emph{Proc. 10th
  {ACM/IEEE} International Symposium on Empirical Software Engineering and
  Measurement, {ESEM}'16}}. \bibinfo{publisher}{{ACM}}.
\newblock


\bibitem[\protect\citeauthoryear{Han, Yu, and Lo}{Han et~al\mbox{.}}{2018}]%
        {DBLP:conf/kbse/HanYL18}
\bibfield{author}{\bibinfo{person}{Xue Han}, \bibinfo{person}{Tingting Yu},
  {and} \bibinfo{person}{David Lo}.} \bibinfo{year}{2018}\natexlab{}.
\newblock \showarticletitle{PerfLearner: learning from bug reports to
  understand and generate performance test frames}. In
  \bibinfo{booktitle}{\emph{Proc. 33rd {ACM/IEEE} International Conference on
  Automated Software Engineering, {ASE} 2018}}. \bibinfo{publisher}{{ACM}}.
\newblock


\bibitem[\protect\citeauthoryear{Herzig, Just, and Zeller}{Herzig
  et~al\mbox{.}}{2014}]%
        {DBLP:conf/icse/HerzigJZ13}
\bibfield{author}{\bibinfo{person}{Kim Herzig}, \bibinfo{person}{Sascha Just},
  {and} \bibinfo{person}{Andreas Zeller}.} \bibinfo{year}{2014}\natexlab{}.
\newblock \showarticletitle{It's not a bug, it's a feature: how
  misclassification impacts bug prediction}. In \bibinfo{booktitle}{\emph{35th
  International Conference on Software Engineering, {ICSE} 2013}}.
  \bibinfo{publisher}{{IEEE} Computer Society}.
\newblock


\bibitem[\protect\citeauthoryear{Humbatova, Jahangirova, Bavota, Riccio,
  Stocco, and Tonella}{Humbatova et~al\mbox{.}}{2020}]%
        {DBLP:conf/icse/HumbatovaJBR0T20}
\bibfield{author}{\bibinfo{person}{Nargiz Humbatova}, \bibinfo{person}{Gunel
  Jahangirova}, \bibinfo{person}{Gabriele Bavota}, \bibinfo{person}{Vincenzo
  Riccio}, \bibinfo{person}{Andrea Stocco}, {and} \bibinfo{person}{Paolo
  Tonella}.} \bibinfo{year}{2020}\natexlab{}.
\newblock \showarticletitle{Taxonomy of real faults in deep learning systems}.
  In \bibinfo{booktitle}{\emph{{ICSE} 2020: 42nd International Conference on
  Software Engineering}}. \bibinfo{publisher}{{ACM}}.
\newblock


\bibitem[\protect\citeauthoryear{Islam, Nguyen, Pan, and Rajan}{Islam
  et~al\mbox{.}}{2019}]%
        {DBLP:conf/sigsoft/IslamNPR19}
\bibfield{author}{\bibinfo{person}{Md~Johirul Islam}, \bibinfo{person}{Giang
  Nguyen}, \bibinfo{person}{Rangeet Pan}, {and} \bibinfo{person}{Hridesh
  Rajan}.} \bibinfo{year}{2019}\natexlab{}.
\newblock \showarticletitle{A comprehensive study on deep learning bug
  characteristics}. In \bibinfo{booktitle}{\emph{Proc. ACM Conference and
  Symposium on the Foundations of Software Engineering, {FSE} 2019}}.
  \bibinfo{publisher}{{ACM}}.
\newblock


\bibitem[\protect\citeauthoryear{Islam, Pan, Nguyen, and Rajan}{Islam
  et~al\mbox{.}}{2020}]%
        {DBLP:conf/icse/IslamPNR20}
\bibfield{author}{\bibinfo{person}{Md~Johirul Islam}, \bibinfo{person}{Rangeet
  Pan}, \bibinfo{person}{Giang Nguyen}, {and} \bibinfo{person}{Hridesh Rajan}.}
  \bibinfo{year}{2020}\natexlab{}.
\newblock \showarticletitle{Repairing deep neural networks: fix patterns and
  challenges}. In \bibinfo{booktitle}{\emph{{ICSE} 2020: 42nd International
  Conference on Software Engineering}}. \bibinfo{publisher}{{ACM}}.
\newblock


\bibitem[\protect\citeauthoryear{Kadam and Bhalerao}{Kadam and
  Bhalerao}{2010}]%
        {kadam2010sample}
\bibfield{author}{\bibinfo{person}{Prashant Kadam} {and}
  \bibinfo{person}{Supriya Bhalerao}.} \bibinfo{year}{2010}\natexlab{}.
\newblock \showarticletitle{Sample size calculation}.
\newblock \bibinfo{journal}{\emph{International journal of Ayurveda research}}
  \bibinfo{volume}{1}, \bibinfo{number}{1} (\bibinfo{year}{2010}).
\newblock


\bibitem[\protect\citeauthoryear{Krizhevsky, Sutskever, and Hinton}{Krizhevsky
  et~al\mbox{.}}{2012}]%
        {DBLP:conf/nips/KrizhevskySH12}
\bibfield{author}{\bibinfo{person}{Alex Krizhevsky}, \bibinfo{person}{Ilya
  Sutskever}, {and} \bibinfo{person}{Geoffrey~E. Hinton}.}
  \bibinfo{year}{2012}\natexlab{}.
\newblock \showarticletitle{ImageNet Classification with Deep Convolutional
  Neural Networks}. In \bibinfo{booktitle}{\emph{26th Annual Conference on
  Neural Information Processing Systems 2012}}.
\newblock


\bibitem[\protect\citeauthoryear{Li, Tan, Wang, Lu, Zhou, and Zhai}{Li
  et~al\mbox{.}}{2006}]%
        {DBLP:conf/asplos/LiTWLZZ06}
\bibfield{author}{\bibinfo{person}{Zhenmin Li}, \bibinfo{person}{Lin Tan},
  \bibinfo{person}{Xuanhui Wang}, \bibinfo{person}{Shan Lu},
  \bibinfo{person}{Yuanyuan Zhou}, {and} \bibinfo{person}{Chengxiang Zhai}.}
  \bibinfo{year}{2006}\natexlab{}.
\newblock \showarticletitle{Have things changed now?: an empirical study of bug
  characteristics in modern open source software}. In
  \bibinfo{booktitle}{\emph{Proc. 1st Workshop on Architectural and System
  Support for Improving Software Dependability, {ASID} 2006}}.
  \bibinfo{publisher}{{ACM}}.
\newblock


\bibitem[\protect\citeauthoryear{Liu, Li, and Chen}{Liu et~al\mbox{.}}{2020}]%
        {DBLP:conf/issta/Liu0020}
\bibfield{author}{\bibinfo{person}{Muyang Liu}, \bibinfo{person}{Ke Li}, {and}
  \bibinfo{person}{Tao Chen}.} \bibinfo{year}{2020}\natexlab{}.
\newblock \showarticletitle{DeepSQLi: deep semantic learning for testing {SQL}
  injection}. In \bibinfo{booktitle}{\emph{{ISSTA} '20: 29th International
  Symposium on Software Testing and Analysis, Virtual Event, USA, July 18-22,
  2020}}. \bibinfo{pages}{286--297}.
\newblock


\bibitem[\protect\citeauthoryear{Liu, Xu, and Cheung}{Liu
  et~al\mbox{.}}{2014}]%
        {DBLP:conf/icse/LiuXC14}
\bibfield{author}{\bibinfo{person}{Yepang Liu}, \bibinfo{person}{Chang Xu},
  {and} \bibinfo{person}{Shing{-}Chi Cheung}.} \bibinfo{year}{2014}\natexlab{}.
\newblock \showarticletitle{Characterizing and detecting performance bugs for
  smartphone applications}. In \bibinfo{booktitle}{\emph{36th International
  Conference on Software Engineering, {ICSE} 2014}}.
  \bibinfo{publisher}{{ACM}}.
\newblock


\bibitem[\protect\citeauthoryear{Mani, Catherine, Sinha, and Dubey}{Mani
  et~al\mbox{.}}{2012}]%
        {DBLP:conf/sigsoft/ManiCSD12}
\bibfield{author}{\bibinfo{person}{Senthil Mani}, \bibinfo{person}{Rose
  Catherine}, \bibinfo{person}{Vibha~Singhal Sinha}, {and}
  \bibinfo{person}{Avinava Dubey}.} \bibinfo{year}{2012}\natexlab{}.
\newblock \showarticletitle{{AUSUM:} approach for unsupervised bug report
  summarization}. In \bibinfo{booktitle}{\emph{20th {ACM} {SIGSOFT} Symposium
  on the Foundations of Software Engineering}}. \bibinfo{publisher}{{ACM}}.
\newblock


\bibitem[\protect\citeauthoryear{McHugh}{McHugh}{2012}]%
        {mchugh2012interrater}
\bibfield{author}{\bibinfo{person}{Mary~L McHugh}.}
  \bibinfo{year}{2012}\natexlab{}.
\newblock \showarticletitle{Interrater reliability: the kappa statistic}.
\newblock \bibinfo{journal}{\emph{Biochemia medica}} \bibinfo{volume}{22},
  \bibinfo{number}{3} (\bibinfo{year}{2012}).
\newblock


\bibitem[\protect\citeauthoryear{Selakovic and Pradel}{Selakovic and
  Pradel}{2017}]%
        {DBLP:conf/se/SelakovicP17}
\bibfield{author}{\bibinfo{person}{Marija Selakovic} {and}
  \bibinfo{person}{Michael Pradel}.} \bibinfo{year}{2017}\natexlab{}.
\newblock \showarticletitle{Performance Issues and Optimizations in JavaScript:
  An Empirical Study}. In \bibinfo{booktitle}{\emph{Software Engineering
  2017}}.
\newblock


\bibitem[\protect\citeauthoryear{Srisakaokul, Wu, Astorga, Alebiosu, and
  Xie}{Srisakaokul et~al\mbox{.}}{2018}]%
        {DBLP:conf/aaai/SrisakaokulWAAX18}
\bibfield{author}{\bibinfo{person}{Siwakorn Srisakaokul},
  \bibinfo{person}{Zhengkai Wu}, \bibinfo{person}{Angello Astorga},
  \bibinfo{person}{Oreoluwa Alebiosu}, {and} \bibinfo{person}{Tao Xie}.}
  \bibinfo{year}{2018}\natexlab{}.
\newblock \showarticletitle{Multiple-Implementation Testing of Supervised
  Learning Software}. In \bibinfo{booktitle}{\emph{The Workshops of the The
  Thirty-Second {AAAI} Conference on Artificial Intelligence, 2018}}
  \emph{(\bibinfo{series}{{AAAI} Workshops}, Vol.~\bibinfo{volume}{{WS-18}})}.
  \bibinfo{publisher}{{AAAI} Press}.
\newblock


\bibitem[\protect\citeauthoryear{Sun, Zhou, Li, Hu, Yang, and Li}{Sun
  et~al\mbox{.}}{2017}]%
        {DBLP:conf/apsec/SunZLHYL17}
\bibfield{author}{\bibinfo{person}{Xiaobing Sun}, \bibinfo{person}{Tianchi
  Zhou}, \bibinfo{person}{Gengjie Li}, \bibinfo{person}{Jiajun Hu},
  \bibinfo{person}{Hui Yang}, {and} \bibinfo{person}{Bin Li}.}
  \bibinfo{year}{2017}\natexlab{}.
\newblock \showarticletitle{An Empirical Study on Real Bugs for Machine
  Learning Programs}. In \bibinfo{booktitle}{\emph{24th Asia-Pacific Software
  Engineering Conference, {APSEC} 2017}}. \bibinfo{publisher}{{IEEE} Computer
  Society}.
\newblock


\bibitem[\protect\citeauthoryear{Tan, Liu, Li, Wang, Zhou, and Zhai}{Tan
  et~al\mbox{.}}{2014}]%
        {DBLP:journals/ese/TanLLWZZ14}
\bibfield{author}{\bibinfo{person}{Lin Tan}, \bibinfo{person}{Chen Liu},
  \bibinfo{person}{Zhenmin Li}, \bibinfo{person}{Xuanhui Wang},
  \bibinfo{person}{Yuanyuan Zhou}, {and} \bibinfo{person}{ChengXiang Zhai}.}
  \bibinfo{year}{2014}\natexlab{}.
\newblock \showarticletitle{Bug characteristics in open source software}.
\newblock \bibinfo{journal}{\emph{Em. Sof. Eng.}} \bibinfo{volume}{19},
  \bibinfo{number}{6} (\bibinfo{year}{2014}).
\newblock


\bibitem[\protect\citeauthoryear{Thung, Wang, Lo, and Jiang}{Thung
  et~al\mbox{.}}{2012}]%
        {DBLP:conf/issre/ThungWLJ12}
\bibfield{author}{\bibinfo{person}{Ferdian Thung}, \bibinfo{person}{Shaowei
  Wang}, \bibinfo{person}{David Lo}, {and} \bibinfo{person}{Lingxiao Jiang}.}
  \bibinfo{year}{2012}\natexlab{}.
\newblock \showarticletitle{An Empirical Study of Bugs in Machine Learning
  Systems}. In \bibinfo{booktitle}{\emph{23rd {IEEE} International Symposium on
  Software Reliability Engineering, {ISSRE} 2012}}. \bibinfo{publisher}{{IEEE}
  Computer Society}.
\newblock


\bibitem[\protect\citeauthoryear{Tian, Lo, Xia, and Sun}{Tian
  et~al\mbox{.}}{2015}]%
        {DBLP:journals/ese/TianLXS15}
\bibfield{author}{\bibinfo{person}{Yuan Tian}, \bibinfo{person}{David Lo},
  \bibinfo{person}{Xin Xia}, {and} \bibinfo{person}{Chengnian Sun}.}
  \bibinfo{year}{2015}\natexlab{}.
\newblock \showarticletitle{Automated prediction of bug report priority using
  multi-factor analysis}.
\newblock \bibinfo{journal}{\emph{Empir. Softw. Eng.}} \bibinfo{volume}{20},
  \bibinfo{number}{5} (\bibinfo{year}{2015}).
\newblock


\bibitem[\protect\citeauthoryear{Tizpaz{-}Niari, Cern{\'{y}}, and
  Trivedi}{Tizpaz{-}Niari et~al\mbox{.}}{2020}]%
        {DBLP:conf/issta/Tizpaz-NiariC020}
\bibfield{author}{\bibinfo{person}{Saeid Tizpaz{-}Niari},
  \bibinfo{person}{Pavol Cern{\'{y}}}, {and} \bibinfo{person}{Ashutosh
  Trivedi}.} \bibinfo{year}{2020}\natexlab{}.
\newblock \showarticletitle{Detecting and understanding real-world differential
  performance bugs in machine learning libraries}. In
  \bibinfo{booktitle}{\emph{29th International Symposium on Software Testing
  and Analysis}}.
\newblock


\bibitem[\protect\citeauthoryear{Uddin, Ghazali, Deris, Naseem, and Shah}{Uddin
  et~al\mbox{.}}{2017}]%
        {DBLP:journals/air/UddinGDNS17}
\bibfield{author}{\bibinfo{person}{Jamal Uddin}, \bibinfo{person}{Rozaida
  Ghazali}, \bibinfo{person}{Mustafa~Mat Deris}, \bibinfo{person}{Rashid
  Naseem}, {and} \bibinfo{person}{Habib Shah}.}
  \bibinfo{year}{2017}\natexlab{}.
\newblock \showarticletitle{A survey on bug prioritization}.
\newblock \bibinfo{journal}{\emph{Artif. Intell. Rev.}} \bibinfo{volume}{47},
  \bibinfo{number}{2} (\bibinfo{year}{2017}).
\newblock


\bibitem[\protect\citeauthoryear{Wen, Liu, Byrne, and Chabbi}{Wen
  et~al\mbox{.}}{2018}]%
        {DBLP:conf/asplos/WenLBC18}
\bibfield{author}{\bibinfo{person}{Shasha Wen}, \bibinfo{person}{Xu Liu},
  \bibinfo{person}{John Byrne}, {and} \bibinfo{person}{Milind Chabbi}.}
  \bibinfo{year}{2018}\natexlab{}.
\newblock \showarticletitle{Watching for Software Inefficiencies with Witch}.
  In \bibinfo{booktitle}{\emph{Proc. of the 23rd International Conference on
  Architectural Support for Programming Languages and Operating Systems}}.
  \bibinfo{pages}{332--347}.
\newblock


\bibitem[\protect\citeauthoryear{Xia, Lo, Wen, Shihab, and Zhou}{Xia
  et~al\mbox{.}}{2014}]%
        {DBLP:conf/csmr/XiaLWSZ14}
\bibfield{author}{\bibinfo{person}{Xin Xia}, \bibinfo{person}{David Lo},
  \bibinfo{person}{Ming Wen}, \bibinfo{person}{Emad Shihab}, {and}
  \bibinfo{person}{Bo Zhou}.} \bibinfo{year}{2014}\natexlab{}.
\newblock \showarticletitle{An empirical study of bug report field
  reassignment}. In \bibinfo{booktitle}{\emph{2014 Software Evolution Week -
  {IEEE} Conference on Software Maintenance, Reengineering, and Reverse
  Engineering}}.
\newblock


\bibitem[\protect\citeauthoryear{Zaman, Adams, and Hassan}{Zaman
  et~al\mbox{.}}{2011}]%
        {DBLP:conf/msr/ZamanAH11}
\bibfield{author}{\bibinfo{person}{Shahed Zaman}, \bibinfo{person}{Bram Adams},
  {and} \bibinfo{person}{Ahmed~E. Hassan}.} \bibinfo{year}{2011}\natexlab{}.
\newblock \showarticletitle{Security versus performance bugs: a case study on
  Firefox}. In \bibinfo{booktitle}{\emph{Proc. 8th International Working
  Conference on Mining Software Repositories, {MSR} 2011}}.
  \bibinfo{publisher}{{ACM}}.
\newblock


\bibitem[\protect\citeauthoryear{Zhang, Wang, Hao, Xie, Zhang, and Mei}{Zhang
  et~al\mbox{.}}{2015}]%
        {DBLP:journals/chinaf/ZhangWHXZM15}
\bibfield{author}{\bibinfo{person}{Jie Zhang}, \bibinfo{person}{Xiaoyin Wang},
  \bibinfo{person}{Dan Hao}, \bibinfo{person}{Bing Xie}, \bibinfo{person}{Lu
  Zhang}, {and} \bibinfo{person}{Hong Mei}.} \bibinfo{year}{2015}\natexlab{}.
\newblock \showarticletitle{A survey on bug-report analysis}.
\newblock \bibinfo{journal}{\emph{Sci. China Inf. Sci.}} \bibinfo{volume}{58},
  \bibinfo{number}{2} (\bibinfo{year}{2015}).
\newblock


\bibitem[\protect\citeauthoryear{Zhang, Xiao, Zhang, Liu, Lin, and Yang}{Zhang
  et~al\mbox{.}}{2020}]%
        {DBLP:conf/icse/ZhangXZLLY20}
\bibfield{author}{\bibinfo{person}{Ru Zhang}, \bibinfo{person}{Wencong Xiao},
  \bibinfo{person}{Hongyu Zhang}, \bibinfo{person}{Yu Liu},
  \bibinfo{person}{Haoxiang Lin}, {and} \bibinfo{person}{Mao Yang}.}
  \bibinfo{year}{2020}\natexlab{}.
\newblock \showarticletitle{An empirical study on program failures of deep
  learning jobs}. In \bibinfo{booktitle}{\emph{{ICSE} 2020: 42nd International
  Conference on Software Engineering}}. \bibinfo{publisher}{{ACM}}.
\newblock


\bibitem[\protect\citeauthoryear{Zhang, Chen, Cheung, Xiong, and Zhang}{Zhang
  et~al\mbox{.}}{2018}]%
        {DBLP:conf/issta/ZhangCCXZ18}
\bibfield{author}{\bibinfo{person}{Yuhao Zhang}, \bibinfo{person}{Yifan Chen},
  \bibinfo{person}{Shing{-}Chi Cheung}, \bibinfo{person}{Yingfei Xiong}, {and}
  \bibinfo{person}{Lu Zhang}.} \bibinfo{year}{2018}\natexlab{}.
\newblock \showarticletitle{An empirical study on TensorFlow program bugs}. In
  \bibinfo{booktitle}{\emph{Proc. 27th {ACM} {SIGSOFT} International Symposium
  on Software Testing and Analysis, {ISSTA} 2018}}. \bibinfo{publisher}{{ACM}}.
\newblock


\bibitem[\protect\citeauthoryear{Zhao, Xiao, Wang, Chen, and Liu}{Zhao
  et~al\mbox{.}}{2019}]%
        {DBLP:conf/icse/Zhao0W0L19}
\bibfield{author}{\bibinfo{person}{Yutong Zhao}, \bibinfo{person}{Lu Xiao},
  \bibinfo{person}{Xiao Wang}, \bibinfo{person}{Bihuan Chen}, {and}
  \bibinfo{person}{Yang Liu}.} \bibinfo{year}{2019}\natexlab{}.
\newblock \showarticletitle{Localized or architectural: an empirical study of
  performance issues dichotomy}. In \bibinfo{booktitle}{\emph{Proc. 41st
  International Conference on Software Engineering}}.
\newblock


\bibitem[\protect\citeauthoryear{Zhao, Xiao, Wang, Sun, Chen, Liu, and
  Bondi}{Zhao et~al\mbox{.}}{2020}]%
        {DBLP:conf/wosp/Zhao0WS0LB20}
\bibfield{author}{\bibinfo{person}{Yutong Zhao}, \bibinfo{person}{Lu Xiao},
  \bibinfo{person}{Xiao Wang}, \bibinfo{person}{Lei Sun},
  \bibinfo{person}{Bihuan Chen}, \bibinfo{person}{Yang Liu}, {and}
  \bibinfo{person}{Andre~B. Bondi}.} \bibinfo{year}{2020}\natexlab{}.
\newblock \showarticletitle{How Are Performance Issues Caused and Resolved?-An
  Empirical Study from a Design Perspective}. In
  \bibinfo{booktitle}{\emph{{ICPE} 2020: {ACM/SPEC} International Conference on
  Performance Engineering}}. \bibinfo{publisher}{{ACM}}.
\newblock


\bibitem[\protect\citeauthoryear{Zhao, Zhang, Shihab, Zou, and Hassan}{Zhao
  et~al\mbox{.}}{2016}]%
        {DBLP:conf/esem/ZhaoZSZH16}
\bibfield{author}{\bibinfo{person}{Yu Zhao}, \bibinfo{person}{Feng Zhang},
  \bibinfo{person}{Emad Shihab}, \bibinfo{person}{Ying Zou}, {and}
  \bibinfo{person}{Ahmed~E. Hassan}.} \bibinfo{year}{2016}\natexlab{}.
\newblock \showarticletitle{How Are Discussions Associated with Bug Reworking?:
  An Empirical Study on Open Source Projects}. In
  \bibinfo{booktitle}{\emph{Proc. 10th {ACM/IEEE} International Symposium on
  Empirical Software Engineering and Measurement, {ESEM} 2016}}.
  \bibinfo{publisher}{{ACM}}.
\newblock


\bibitem[\protect\citeauthoryear{Zimmermann, Premraj, Bettenburg, Just,
  Schr{\"{o}}ter, and Weiss}{Zimmermann et~al\mbox{.}}{2010}]%
        {DBLP:journals/tse/ZimmermannPBJSW10}
\bibfield{author}{\bibinfo{person}{Thomas Zimmermann}, \bibinfo{person}{Rahul
  Premraj}, \bibinfo{person}{Nicolas Bettenburg}, \bibinfo{person}{Sascha
  Just}, \bibinfo{person}{Adrian Schr{\"{o}}ter}, {and}
  \bibinfo{person}{Cathrin Weiss}.} \bibinfo{year}{2010}\natexlab{}.
\newblock \showarticletitle{What Makes a Good Bug Report?}
\newblock \bibinfo{journal}{\emph{{IEEE} Trans. Software Eng.}}
  \bibinfo{volume}{36}, \bibinfo{number}{5} (\bibinfo{year}{2010}).
\newblock


\end{thebibliography}
